\documentclass[aps,prd,onecolumn,nofootinbib,superscriptaddress]{revtex4-2}
\pdfoutput=1
\usepackage{float}
\usepackage{graphicx}
\usepackage{amsmath}
\usepackage{enumitem}
\usepackage{empheq}
\PassOptionsToPackage{centertags}{amsmath}
\usepackage{amsfonts}
\usepackage{amsmath}
\usepackage{xcolor}
\definecolor{mylightblue}{RGB}{100, 149, 237}  
\definecolor{skyblue}{RGB}{135, 206, 235}
\usepackage{amssymb}
\usepackage{amssymb,ulem}
\makeatletter
\renewcommand{\@makefnmark}{%
  \hbox{\@textsuperscript{\normalfont\textcolor{blue}{\@thefnmark}}}
}
\usepackage{dcolumn}
\usepackage{amsmath,amssymb,mathrsfs}
\usepackage{slashed}
\usepackage{subfigure}
\usepackage{multirow}
\usepackage{csquotes}

\usepackage{MnSymbol,wasysym}
\usepackage{braket,diagbox}
\usepackage{eurosym}
\usepackage[usenames,dvipsnames,svgnames]{xcolor}

\newcommand{\RNum}[1]{\uppercase\expandafter{\romannumeral #1\relax}}
\usepackage[colorlinks=true,linkcolor=blue,urlcolor=blue,filecolor=black,citecolor=red,
pdfstartview=FitV,pdftitle={},pdfsubject={},pdfkeywords={},pdfpagemode=None,bookmarksopen=true]{hyperref}

\usepackage{pifont}
\usepackage{booktabs}

\begin{document}
\baselineskip=0.5 cm

\title{Soft cutoffs in the covariant phase space of dynamical reference frames}

\author{Kang Liu}
\email{liukanyang@sina.com}
\affiliation{\baselineskip=0.4cm Center for Gravitation and Cosmology, College of Physical Science and Technology, Yangzhou University, Yangzhou, 225009, China}

\author{Wei Guo}
\email{wguo@cwnu.edu.cn}
\affiliation{\baselineskip=0.4cm School of Physics and Astronomy, China West Normal University, Nanchong 637009, China}

\author{Xiao-Mei Kuang}
\email{xmeikuang@yzu.edu.cn}
\affiliation{\baselineskip=0.4cm Center for Gravitation and Cosmology, College of Physical Science and Technology, Yangzhou University, Yangzhou, 225009, China}

\setcounter{tocdepth}{2}

\begin{abstract}
	\baselineskip=0.45 cm
	 We construct covariant theories which incorporate fluctuating boundaries and soft cutoffs by introducing dynamical reference frames (DRFs). This framework generalizes the covariant action from a hard-cutoff to a soft-cutoff formulation, utilizing smearing functions and their corresponding operator expansions. This generalization initially leads to a loss of diffeomorphism covariance, which is recovered solely by restricting the DRFs, along with both their associated and linear MCFs, to specific forms, and by imposing suitable boundary conditions on the smearing functions. Satisfying these conditions restores covariance in relational spacetime, thereby enabling the consistent definition of subsystems. Within the covariant phase space formalism, we derive the charges of the soft-cutoff theory while explicitly addressing the inherent ambiguities arising from the boundary Lagrangian. We demonstrate that introducing an additional pointwise dependence is essential to resolve these ambiguities and ensure the integrability of the charges, even under fluctuating boundary conditions. Finally, in the context of General Relativity (GR), we establish the conditions under which holographic renormalization results at the asymptotic boundary coincide with the Noether charges derived from our soft-cutoff procedure.
\end{abstract}
\maketitle
\tableofcontents

\section{Introduction}
Defining local subsystems remains a fundamental challenge in gravity. Unlike local quantum field theories, where the Hilbert space naturally factorizes across spatial subregions, that of diffeomorphism-invariant theories lacks a simple tensor product decomposition \cite{Donnelly:2014fua}.
A cornerstone of the canonical formulation of gravity \cite{Dirac:1958sc} is diffeomorphism invariance, which dictates that local fields evaluated at fixed coordinate points are not gauge-invariant and thus cannot serve as physical observables \cite{Bergmann:1961wa,DeWitt:1962cg}. Consequently, rigorous no-go theorems establish the absence of local observables in universes with compact spatial slices \cite{Torre:1993fq}. This constraint implies that physical quantities must be non-local or relational, exhibiting inherent correlations across the system \cite{Rovelli:2001bz,Dittrich:2005kc}.

The covariant phase space formalism provides a standard mathematical framework for addressing these issues \cite{Iyer:1994ys,lee1990local,Wald:1999wa,Barnich:2001jy,Harlow:2019yfa}. While this formalism has successfully derived the first law of black hole mechanics and quasi-local energy expressions such as the Brown-York stress tensor \cite{Komar1959CovariantCL,Brown:1992br}, and its traditional application often relies on fixed boundary conditions. However, the integrability of charges in this scenario is often obstructed when the theory involves variable coupling constants or non-stationary horizons \cite{PhysRevD.7.2333,Wald:1999wa,Hollands:2012sf}.  This approach effectively promotes the couplings to the conserved charges of a new global gauge symmetry \cite{Chernyavsky:2017xwm,Hajian:2021hje,Hajian:2023bhq}, ensuring their variations are well-defined phase space operations and restoring the integrability of the first law \cite{Smarr:1972kt}. Conversely, for dynamical black holes where symplectic flux across the horizon prevents a canonical charge definition\cite{Rignon-Bret:2023fjq,Barnich:2001jy}, the integrability is recovered by identifying a specific boundary correction term for the pre-symplectic potential \cite{Gao:2001ut}.

The foundation of edge modes lies in the study of topological phases where boundary excitations emerge in the quantum Hall effect \cite{Halperin:1981ug,Wen:1992vi}. In these systems the bulk Chern-Simons theory requires gapless edge currents to preserve gauge invariance at the spacetime boundary \cite{Cordes:1994fc,Witten:2009at}. Drawing a direct parallel to this topological mechanism seminal studies established that diffeomorphism invariance in gravity similarly implies the existence of edge states on spatial boundaries \cite{Balachandran:1994up,Balachandran:1995qa}. It was demonstrated that degrees of freedom at the boundary are necessary to define a non-degenerate symplectic structure \cite{Crnkovic:1986ex,kijowski1976canonical}, and gauge transformations supported on the boundary become spacetime symmetries generated by non-vanishing charges \cite{Regge:1974zd,Geiller:2017xad,Liu:2025jvj}. Subsequent research utilized edge modes to resolve the factorization problem in calculating entanglement entropy for local subsystems \cite{Donnelly:2016auv,Geiller:2017xad,Speranza:2017gxd,Geiller:2017whh,Freidel:2020xyx,Freidel:2020svx}. Just as edge states in the Hall effect encode topological data \cite{Carlip:1994gy}, the gravitational edge modes restore gauge invariance at the entangling surface \cite{Donnelly:2014fua,Buividovich:2008gq} and permit the definition of an extended Hilbert space \cite{Ghosh:2015iwa,Aoki:2015bsa}. This construction provides a rigorous derivation for the area law of black hole entropy \cite{Sorkin:1984kjy,PhysRevD.34.373} and resolves ambiguities in the definition of pre-symplectic potentials \cite{Sorkin:1984kjy,Jacobson:1993vj}.

The proposal of dressed observers posits that in gravitational theories, ensuring the diffeomorphism invariance of local operators requires attaching a gravitational dressing that extends to the asymptotic boundary \cite{Heemskerk:2012np,Kabat:2013wga,Donnelly:2016rvo,Carrozza:2022xut,Carrozza:2021gju}. To bypass the inability to factorize the Hilbert space into subsystems, \cite{Donnelly:2017jcd,Giddings:2019hjc} proposes a weakened `` split structure" framework: instead of algebraically isolating operators, it identifies specific sets of Hilbert space states where different internal matter configurations remain completely indistinguishable to any observable outside a thickened boundary. This long-range dependence renders the operators kinematically non-local and leads to non-vanishing commutators for spacelike separated dressed operators due to the overlap of their gravitational tails, manifesting as a violation of microcausality within a perturbative context \cite{Donnelly:2015hta,Giddings:2015lla,Giddings:2005id,Donnelly:2017jcd}.
To address this issue, physicists consider the DRF, which introduces a relational description based on physical field mappings, elevating coordinates from spurious background labels to the locations defined by the fields, thereby establishing relational locality \cite{Goeller:2022rsx,Vanrietvelde:2018pgb,Hohn:2018toe,Castro-Ruiz:2025yvi,Araujo-Regado:2025ejs}. Building on this, by applying a smearing procedure to relational observables, one can construct operators with compact relational support. This further eliminates the singularities associated with line-supported dressings and allows observables in single-integral form to approximately recover local physics in specific states, thereby establishing a well-defined local subalgebra structure within the relational algebra \cite{Dittrich:2005kc}. Furthermore, through non-perturbative analysis of frame-dressed fields in relational space using Peierls brackets \cite{10.1098/rspa.1952.0158}, it was demonstrated that as long as the relational supports of two observables remain spacelike separated under the field-dependent metric, their commutators strictly vanish, thus successfully restoring bulk microcausality from a gauge-invariant relational perspective \cite{Carrozza:2022xut}.

Moreover, the resolution of the problem of time and the vanishing of local observables relies on  DRFs, which are regarded  as the edge modes \cite{Riello:2020zbk,Wieland:2020gno}, and physicalize the coordinate system \cite{Tambornino:2011vg,Hohn:2018iwn}.  This framework fundamentally distinguishes between internal frames which restore gauge invariance in Yang-Mills theories \cite{Marsh:2016hdj,Gomes:2016mwl,Gu:2009jh} and spacetime frames which fix diffeomorphism invariance in gravity by anchoring fields to matter \cite{Brown:1994py,Kuchar:1991qf,Isham:1992ms}. Due to the non-decoupling of boundary modes in such non-Abelian theories, their edge mode entanglement entropy reduces to the thermodynamic entropy of a ghost on a codimension-two hypersurface \cite{Ball:2024gti,Ball:2024xhf}. The geometry of these frame transformations is rigorously captured by the MCFs, which defines a connection to distinguish physical variations from gauge redundancies \cite{Henneaux:1992ig,Zuckerman:1986vzu,Khavkine:2014kya} and also induces the extended phase space \cite{Ciambelli:2021nmv,Freidel:2021dxw} as embedding fields.

However, while the introducing of DRFs addresses the locality problem of gravity to some extent, potential unresolved issues remain within this framework. On the one hand, while the path integral measure is divided by the gauge group volume to remove redundancy in unbounded manifolds \cite{Witten:1991we,Donnelly:2016auv}, the introducing of a boundary promotes ``large" gauge transformations into dynamical boundary fields. Consequently, the edge partition function becomes an integral over these boundary degrees of freedom, necessitating precise measure handling to account for bulk versus boundary group volumes \cite{Blommaert:2018oue}. A natural question is whether similar issues extend to DRFs associated with diffeomorphisms, specifically, whether DRFs can influence the integration measure of the covariant theory and thereby affect the boundary dynamics.

On the other hand, to fully capture the dynamics of the boundaries, it is necessary to extend this framework to include fluctuating boundaries by introducing Heaviside step function and Dirac delta function in the method of the covariant phase space \cite{Adami:2024gdx,Golshani:2024fry}. Allowing the boundary to wiggle introduces new dynamical degrees of freedom and modifies the symplectic structure, a feature that has been shown to be crucial for understanding the ``wiggling throat'' of extremal black holes and the integrability of surface charges in the presence of radiation \cite{Compere:2015bca,Hollands:2024vbe}. In the ``nearly-$\rm{AdS_2}$" formalism, the rigid asymptotic boundary is replaced by a fluctuating curve, effectively elevating the boundary geometry to a dynamical degree of freedom \cite{Maldacena:2016upp}. This factorization captures the ``would-be gauge modes" that arise from broken radial diffeomorphisms \cite{Joung:2023doq,Carlip:2005tz,Liu:2025cjl}. Consequently, the path integral over these boundary fluctuations naturally reproduces the Schwarzian effective action and provides a precise derivation of its measure \cite{Saad:2019lba,Cotler:2018zff}.

Thus, to address the aforementioned issues, in this work we utilize DRFs to naturally introduce fields describing fluctuating boundaries in covariant theories. Analogous to the factors in Janus \cite{Freedman:2003ax,Bak:2011ga} or domain wall \cite{Emparan:2000fn,Melfo:2002wd} solutions, we aim to characterize the boundary using additional fields derived from smooth function generated by the DRFs that recovers the expansion of the Dirac delta function and its derivatives in a specific limit. Our proposal enriches the boundary structure, allowing us to derive its physical implications within the covariant phase space formalism.

This paper is organized as follows. In Section \ref{sec: MCF in relational spacetime}, we briefly review DRFs and present the explicit expressions for the MCFs. In Section \ref{sec:Covariant theories with fluctuating and smearing boundaries}, we introduce a specific DRF and its corresponding smearing function constructed from smooth step functions involving a thickness field. We subsequently derive the associated Dirac delta function expansion and the boundary limit of the MCFs. In Section \ref{sec:Covariant phase space with the fluctuating and smearing boundary}, we compute the symplectic potential and symplectic form for fluctuating boundaries from the soft-cutoff action, addressing the inherent ambiguities. Within the covariant phase space formalism, we demonstrate that the field-dependent part of the thickness field $\epsilon$ can be absorbed into the hypersurface displacement term, allowing us to derive conservation laws and integrable charges. In Section \ref{sec:Noether charge at the asymptotic boundary of general relativity in the FG gauge}, we illustrate the physical significance of the integrable charges within the soft-cutoff action, using holographic renormalization of GR at asymptotic boundaries as an example. Section \ref{sec:conclusions} contains our conclusions and discussion. Additionally, three appendices are included to detail the conventions and calculations.

\section{Dynamical reference frame and Maurer-Cartan form in relational spacetime}
\label{sec: MCF in relational spacetime}

We localize a subregion of spacetime using a DRF, defined as a smooth map $U: \mathcal{M} \rightarrow \mathfrak{m}$ from the spacetime manifold $\mathcal{M}$ with coordinates $x^\mu$ to the relational spacetime $\mathfrak{m}$ with coordinates $y^A$ \cite{Goeller:2022rsx,Carrozza:2021gju,Carrozza:2022xut}. Consequently, the spacetime subregion $M \subset \mathcal{M}$ is defined as the preimage of a region $m \subset \mathfrak{m}$, i.e., $M := U^{-1}(m)$. The coordinate relationship is given by
\begin{align}
y^A = U^A(x^\mu, \Phi) \quad \Longrightarrow \quad x^\mu = (U^{-1})^\mu(y^A; \Phi),
\end{align}
where $\Phi$ denotes the dynamical fields defining the frame. To account for the field-dependence of the frame, we introduce the MCF $\chi$, a one-form in field space defined as:
\begin{equation}\label{Maurer-Cartan}
\chi := \delta U^{-1} \circ U.
\end{equation}
 This form captures the frame variation mapped back to spacetime coordinates. To see this, given that the coordinates $x^\mu$ are independent of the fields, applying the variational chain rule to the frame-dressed identity $x^\mu = (U^{-1})^\mu(U^A(x); \Phi)$ \cite{Donnelly:2016rvo} yields the component expression for $\chi$:
\begin{align}\label{MCF in component}
\delta[{x}^\mu(U^A(x);\Phi)]=0 \quad \Longrightarrow \quad \chi^\mu(x):=\delta x^\mu|_{y} = -\frac{\partial{x}^\mu}{\partial y^A}\delta U^A(x),
\end{align}
where the variational chain rule $\delta[f(g)]={(\delta f)(g)}+{\frac{\partial f}{\partial g}\delta g}$ is used\footnote{This chain rule is illustrated by an example \cite{1997Cft/}. Let us assume an infinitesimal coordinate transformation, such as a conformal transformation $x^\mu\to x^{\prime\mu}=x^\mu+s^\mu(x)$. Under such a transformation, the field $\Phi(x)$ becomes $\Phi'(x')$. The total variation thus consists of the modification of the functional form and the variation arising from the displacement of the coordinate points: $\delta[\Phi(x)]=\delta\Phi(x)-s^\mu\partial_\mu\Phi(x)$.}.

Consider a covariant theory whose frame-dressed action associated with the DRF takes the form
\begin{equation}
S_D = \int_\mathfrak{m} U^* \mathcal{L}(\Phi),
\label{eq:23}
\end{equation}
where $U^*$ and its inverse $U_*: \mathfrak{m} \to \mathcal{M}$ denote the induced pushforward and pullback, respectively. The Lagrangian $D$-form is $\mathcal{L} = L\,\mathrm{d}^Dx$, with $\mathrm{d}^Dx := \mathrm{d}x^1 \wedge \cdots \wedge \mathrm{d}x^D$.

We now present the variational rules for the dressed action $S_D$. Considering an arbitrary differential form $\alpha$ on the spacetime manifold, the variation of its pushforward satisfies
\begin{equation}\label{variation rule}
	\delta(U^*\alpha) = U^*(\delta_{\chi}\alpha),
\end{equation}
where $\delta_{\chi} := \delta + \mathcal{L}_{\chi}$ is the field-space covariant variation, with $\mathcal{L}_{\chi}$ denoting the Lie derivative along $\chi$. Applying this formalism to the Lagrangian ${\cal L}$, the fundamental variational identity on relational spacetime is given by\begin{equation}\label{variation UL}\delta (U^*{\cal L}) = U^*(E_\Phi\delta\Phi) + {\rm d}U^*[\Theta_{\chi}].
\end{equation}
Here, $\delta (U^*{\cal L})$ denotes the variation of the frame-dressed Lagrangian. The boundary term defines the extended pre-symplectic potential $\Theta_{\chi}$ associated with the invariant Lagrangian \cite{Speranza:2017gxd,Ciambelli:2021nmv,Freidel:2021dxw}:
\begin{equation}
\Theta_{\chi} := \Theta_{LW} + i_\chi{\cal L},
\label{extended symplectic potential}
\end{equation}
where $\Theta_{LW}$ is the Lee-Wald pre-symplectic potential \cite{lee1990local}, and $i_\chi{\cal L}$ denotes the contraction of the Lagrangian along the vector field $\chi$. {In Appendix \ref{app:Proofs for equations regarding DRFs}, we give a proof on the  relation \eqref{variation rule}, and also list key formulas for differential forms and present \eqref{variation UL} in explicit component form, which helps illustrate the definition \eqref{Maurer-Cartan} and the Jacobian variation rules.}

\section{Covariant theories with fluctuating and smearing boundaries}
\label{sec:Covariant theories with fluctuating and smearing boundaries}
With these preparations in place, we proceed to introduce boundary degrees of freedom via DRF-induced transformations, effectively characterizing the fluctuations of boundaries subject to soft cutoffs.

Let $b$ and $c$ be coordinates on $\mathcal{M}$ restricted to the intervals $[b_L, b_R]$ and $[c_D, c_U]$. We denote the remaining coordinates by $\hat{x}$ and the coordinates excluding $b$ by $\bar{x}$. With the Lagrangian expressed as a spacetime $D$-form ${\cal L}=\sqrt{-g}L{\rm d}b\wedge{\rm d}c \wedge{\rm d}\hat{x}$, the action for this covariant theory can be cast into the following form using rectangular functions $H(b_L,b_R)$ and $H(c_D,c_U)$:
\begin{align}\label{action begin1}
	{S} = \int_{{b_L}}^{ b_R }\int_{{c_D}}^{ c_U } {\cal L}  = \int_{ - \infty }^{ + \infty } {\cal L} H(b_L,b_R)H(c_D,c_U),
\end{align}
where we have omitted the integral over $\hat{x}$. The rectangular function $H$ is defined as
\begin{align}
	H({x_0},x_1)=
	\begin{cases}
		{}\:1 &  x_0\le x\le x_1 \\
		{}\:0 & \text{otherwise}
	\end{cases},
	\label{step function H}
\end{align}
where $x \in \{b, c\}$, and we have extended the ranges of both $b$ and $c$  to $(-\infty, \infty)$. We mention that the relationship between the variation of $H({x_0},x_1)$ and the Dirac delta function $\Delta(x_{0,1})$ is given by $\delta H(x_0,x_1)=\Delta (x_0)\delta x_0+\Delta(x_1)\delta x_1$ and ${\rm d}H(x_0,x_1)=\Delta (x_0){\rm d} x_0+\Delta(x_1){\rm d} x_1$, which will be used to calculate the symplectic form.

We define the relational coordinates $\mathcal{B}, \mathcal{C}$ via the DRFs:
\begin{align}\label{relational coordinates}
    U_{b,c}:b,c\to {\cal B}= U_b(b,\epsilon),\:\:{\cal C}= U_c(c,\epsilon).
\end{align}
Crucially, the boundaries $b_{L,R}$ and $c_{D,U}$ can be considered to be field-dependent. They are allowed to fluctuate, meaning that neither $\delta b_L$ nor $\delta b_R$ vanishes. Subsequently, we introduce  fields $\alpha_{L,R}$ and $\beta_{C,D}$ to characterize these fluctuations, which satisfy $\delta b_L=b_L-\alpha_L$, $\delta b_R=\alpha_R-b_R$, $\delta c_D=c_D-\beta_L$ and $\delta c_U=\beta_R-c_U$ within the intervals $ [b_L, b_R] \subset [\alpha_L, \alpha_R]$ and $[c_D, c_U] \subset [\beta_D, \beta_U]$. To further characterize the magnitude of these fluctuations, we can introduce an additional thickness field  $\epsilon$ satisfying the orders $\delta b_{L,R} \sim {\cal O} (\epsilon)$ and $\delta c_{D,U}\sim{\cal O} (\epsilon)$. The fields $b_{L,R}$ and $\epsilon$ now define the field space coordinates $Z^{I}=(\epsilon,b_{L},b_{R})$, in whose basis the variation $\delta$ can be expanded:
\begin{subequations}\label{delta basis}
\begin{align}
    &\delta=\sum_I\delta Z^I\wedge\frac{\partial}{\partial Z^I}=\delta\epsilon\wedge\frac{\partial}{\partial\epsilon}+\delta b_L\wedge\frac{\partial}{\partial b_L}+\delta b_R\wedge\frac{\partial}{\partial b_R},\\
    \Longrightarrow& \delta{\cal B}=\delta U_b=\delta\epsilon\frac{\partial U_b}{\partial\epsilon}+\delta b_L\frac{\partial U_b}{\partial b_L}+\delta b_R\frac{\partial U_b}{\partial b_R},\quad\delta{\cal C}=\delta U_c=\delta\epsilon\frac{\partial U_c}{\partial\epsilon}+\delta c_D\frac{\partial U_c}{\partial c_D}+\delta c_U \frac{\partial U_c}{\partial c_D}\label{dH},
    \end{align}
\end{subequations}
which implies that the MCF is constructed to compensate for variations in all scalar fields ($\epsilon, b_{L,R}, c_{D,U}$). We now give the DRFs a concrete form in \eqref{relational coordinates} by specifying $U_{b,c}$, which are defined by the smooth step functions $G_{L,R}(x)$ with the following properties
\begin{align}\label{asympotic properties}
\lim_{u_{L,R}\to-\infty}G_{L,R}(u_{L,R})=0,\quad\lim_{u_{L,R}\to\infty}G_{L,R}(u_{L,R})=1,
\end{align}
with $u_L=\frac{b-b_L}{\epsilon}$ and $ u_R=\frac{b-b_R}{\epsilon}$. For simplicity, we focus on introducing the definition of $U_b$, such that the expression for $U_c$ can be obtained simply by performing a variable substitution on $U_b$. By introducing $U_{b}$ as
\begin{equation}\label{U_distri}
	\mathcal{B} = U_b(b)=\int^{b}_{b_0}\left[G_L\left(u_L\right)-G_R\left(u_R\right)\right]{\rm d}\tau,
\end{equation}
with $b_L< b_0< b_R$, we can explicitly calculate the MCF $\chi$, which serves as a compensation field to realize the relations
\begin{align}\label{covariant variantion of B}
    \delta_\chi{\cal B}=\delta{\cal B}+\chi^\mu\partial_\mu {\cal B}=0.
\end{align}
Consequently, $\delta {\cal B}$ represents the variations of these defining functions in the solution space.

  To extend the covariant theory, we introduce a smearing function ${\cal H}_b={\cal H}_b(u_{L,R})$. Based on \eqref{U_distri}, the corresponding smearing function is defined as
\begin{align}\label{smearing function}
\mathcal{H}_b(u_{L,R}):=G_L(u_L)-G_R(u_R).
\end{align}
We can check that, in the limit of $\epsilon\to 0$, the smearing function may fail to be equivalent to the product of the rectangular functions for the two boundaries mentioned in \cite{Adami:2024gdx} in the hard cutoffs, since $G_{L,R}$ are unassigned at the boundaries $b_{L,R}$.

\begin{figure}[ht!]
    \centering
    \includegraphics[width=0.8\linewidth]{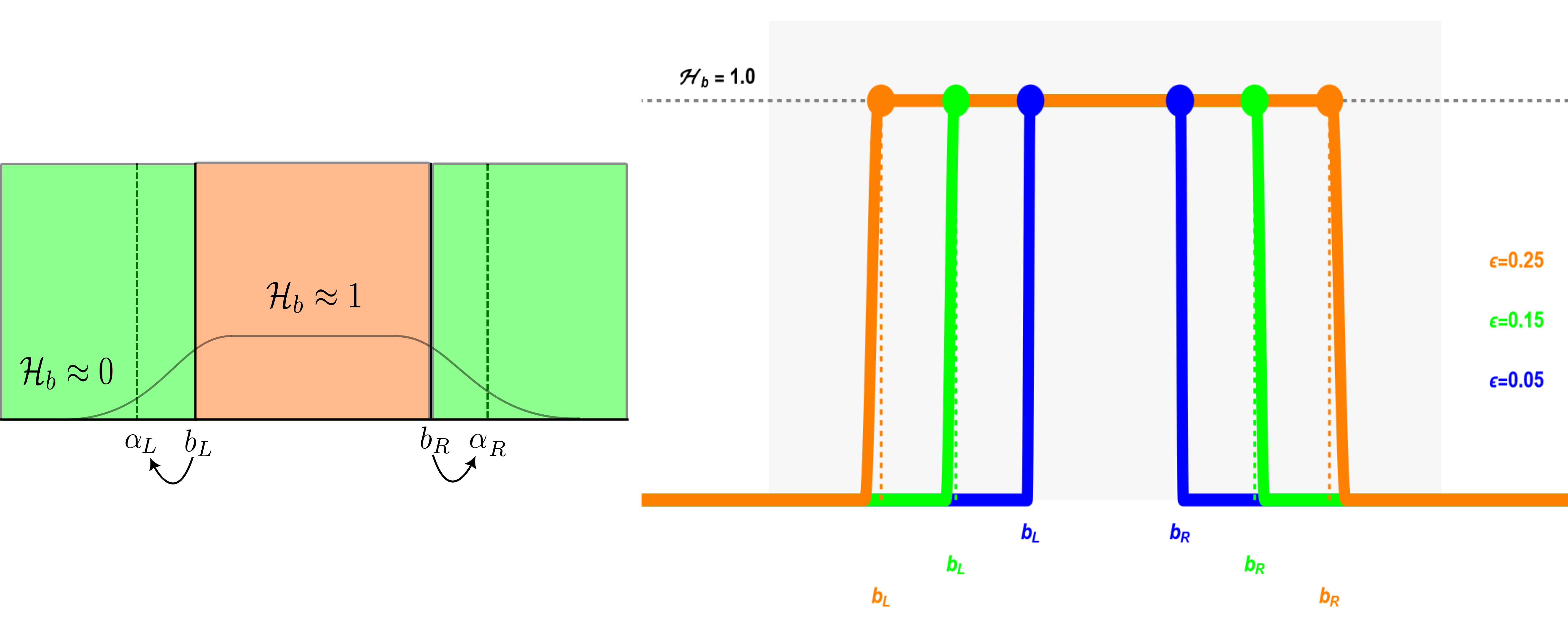}
    \caption{\textbf{Left panel:} The brown region represents the range of $b$ in the original spacetime, while the green region denotes the relational spacetime extended via the boundary mapping induced by the DRFs. Within the interval $[b_L, b_R]$, $\mathcal{H}_b \approx 1$, whereas outside this region, it approaches the small value set by the thickness field. The fluctuating boundaries in spacetime are characterized by the deviation between $b_{L,R}$ and $\alpha_{L,R}$. \textbf{Right panel:} Plots of $\mathcal{H}_b$ evaluated from equation \eqref{example_G} are displayed for various values of $\epsilon$, obtained by explicitly setting $F=0$.}
    \label{fig1}
\end{figure}

The forms of $G_{L,R}$ in \eqref{U_distri} stem from the requirement that the smearing function is generated by a smooth DRF. Integrating this function over the domain $[\alpha_L, \alpha_R]$ then yields the extended action\footnote{Note that the boundaries  $\alpha_{L,R}$ of the extended action can be finite or infinite. In the infinite limit, they are consistent with the upper and lower integration limits in \eqref{action begin1}.}:
 \begin{align}\label{extended action}
     S_\epsilon=\int_{\alpha_L}^{\alpha_R}{\cal H}_b{\cal L},
 \end{align}
  which recovers the original action $S$ up to a correction of order ${\cal O}(\epsilon)$ by
\begin{align}
\begin{aligned}\label{difference action}
S-S_{\epsilon}={\int_{b_{R}}^{\alpha_{R}}\mathcal{L}}+{\int_{\alpha_{L}}^{b_L}\mathcal{L}}+{\int_{b_{L}}^{b_{R}}\left(1-\mathcal{H}_{b}\right)\mathcal{L}}\sim{\cal O(\epsilon)}.
\end{aligned}
 \end{align}
Here we consider $\mathcal{H}_b$ to be of order $\mathcal{O}(\epsilon)$ in the intervals $[\alpha_L, b_L)$ and $(b_R, \alpha_R]$, and $\mathcal{H}_b \approx 1$ in the interval $[b_L, b_R]$, so that the above relation holds, assuming that $\mathcal{L}$ is not of order $\mathcal{O}(1/\epsilon)$ within $(\alpha_L, \alpha_R)$. We show the schematic behavior of $\mathcal{H}_b$ as a function of $b$ in the left panel of FIG. \ref{fig1}. We can see that this choice for $\mathcal{H}_b$ is essentially a generalization of the rectangular functions in \eqref{action begin1}.

Further generalizing the Lagrangian density in \eqref{extended action} yields a form that includes a boundary term $\ell$, shifting the density as $\sqrt{-g}L \, {\mathrm{d}}^D x \to (\sqrt{-g}L + \partial_\mu \ell^\mu) \, {\mathrm{d}}^D x$. This shift reflects the inherent ambiguities discussed in \cite{Wald:1999wa, Speranza:2022lxr}. Furthermore, we can consider the covariant variation of $\ell^\mu$. Motivated by the Gibbons-Hawking term, we take $\ell^\mu$ to be a function of the first and second derivatives of the normal vector. Since $\partial_\mu \mathcal{B}$ acts as the boundary normal in the following section, we have $\ell^\mu=\ell^\mu(\partial_\alpha\mathcal{B},\partial_\alpha\partial_\beta\mathcal{B})$. It can then be shown that the action of $\delta_\chi$ on $\ell^\mu$ yields flowing definition
\begin{align} \label{dxl}
    \mathcal{A}^\mu:=\delta_\chi \ell^\mu=\delta_\chi[\ell^\mu]+\frac{\partial\ell^\mu}{\partial(\partial_\alpha\mathcal{B})}\delta_\chi(\partial_\alpha\mathcal{B})+\frac{\partial\ell^\mu}{\partial(\partial_\alpha\partial_\beta\mathcal{B})}\delta_\chi(\partial_\alpha\partial_\beta\mathcal{B})=\delta_\chi[\ell^\mu]-\frac{\partial\ell^\mu}{\partial(\partial_\alpha\partial_\beta\mathcal{B})}(\partial_\alpha\partial_\beta\chi^\nu)\partial_\nu\mathcal{B},
\end{align}
which incorporates relations $\mathcal{L}_{\chi}(\partial_\alpha\mathcal{B})=\chi^\lambda\partial_\lambda(\partial_\alpha\mathcal{B})+\partial_\alpha\chi^\lambda\partial_\lambda\mathcal{B}$ and $\mathcal{L}_{\chi}(\partial_\alpha\partial_\beta\mathcal{B})=\chi^\lambda\partial_\lambda(\partial_\alpha\partial_\beta\mathcal{B})+\partial_\alpha\chi^\lambda\partial_\lambda\partial_\beta\mathcal{B}+\partial_\beta\chi^\lambda\partial_\alpha\partial_\lambda\mathcal{B}$, and $\delta_\chi[\ell^\mu]$ denotes the variation of $\ell^\mu$ with respect to all fields other than the DRF. Here, when $\mathcal{A}^\mu \neq 0$, the portion independent of $\delta_\chi[\ell^\mu]$ arises from the fact that $\mathcal{A}$ cannot be expressed entirely as a function of the DRF. Relation \eqref{dxl} will be used in Section \ref{subsec:Symplectic forms and integrable charges} to calculate the contraction of the symplectic form.

According to the action extension in \eqref{extended action}, this leads to
\begin{align}\label{ambigui extension}
    S_\epsilon=\int_{\alpha_L}^{\alpha_R}({\cal H}_b{\cal L}-\partial_\mu{\cal H}_b\ell^\mu).
\end{align}
Requiring a well-defined boundary variation, such as the Gibbons-Hawking term in general relativity \cite{gibbons1977action}, allows the extended action \eqref{ambigui extension} to take a more general form.

However, this extended Lagrangian \eqref{extended action}, which explicitly depends on the coordinates $b$, $c$ and the newly introduced $\epsilon$-field , clearly violates the covariance of the theory. This makes it difficult to define any gauge-invariant subsystems. \textit{Our aim is to simultaneously eliminate the explicit dependence on $\epsilon$ and the coordinates $b$ and $c$ by constructing suitable DRFs in relational spacetime, thereby making it possible to define such subsystems.}.

 Motivated by the structure of \eqref{action begin1}, we generalize the action by incorporating DRFs. The Lagrangian in the relational spacetime takes the form $ \widetilde{\cal L}=(\sqrt{-g}L+\partial_\mu\ell^\mu){\rm d}{\cal B}\wedge{\rm d}{\cal C} \wedge{\rm d}^{D-2}{\hat{x}}$.  Then we can express the action \eqref{extended action} on the relational spacetime as
\begin{align}\label{relational_action}
S_\epsilon=\int_{\mathcal{B}_L}^{\mathcal{B}_R} \widetilde{\mathcal{L}} = \int_{b_L}^{b_R}\frac{{\rm d}\mathcal{B}}{{\rm d} b} \frac{{\rm d}\mathcal{C}}{{\rm d} c}(\sqrt{-g} L+\partial_\mu\ell^\mu){\rm d}b\wedge{\rm d}c \wedge{\rm d}^{D-2}{\hat{x}},
\end{align}
with the relational boundaries ${\cal B}_{L,R}={\cal B}(b=b_{L,R})$.

Now the DRF $\mathcal{B}$ acts as the generator of $\mathcal{H}_b$, given by $\frac{\mathrm{d}\mathcal{B}}{\mathrm{d}b} = \mathcal{H}_b$, which recovers the form of \eqref{ambigui extension}. Moreover, to distinguish them from the fluctuating boundaries in spacetime, the above results require the adoption of hard-cutoff boundary conditions:
\begin{align}\label{relational boundary condition}
    \delta{\cal B}_{L,R}=0.
\end{align}
These indicate that the soft cutoffs introduced in spacetime can be regarded to be equivalent to hard cutoffs in relational spacetime. At this point, the Lagrangian in relational spacetime exhibits no explicit coordinate dependence in order to preserve covariance, which provides the motivation for defining the smearing function in \eqref{relational_action} via factors $\frac{\mathrm{d}b}{\mathrm{d}\mathcal{B}}$.

For the sake of feasibility,  we provide a simple example of the smooth step functions:
\begin{subequations}\label{example_G}
\begin{align}
&G_L(u_L,\bar x):=S(u_L+1,\bar x), \\
&G_R(u_R,\bar x):=S(u_R,\bar x),
\end{align}
\end{subequations}
where $S(u , \bar{x})=\frac{\phi(u)}{\phi(u)+e^{F(u, \bar{x})} \phi(1-u)}$, with $\phi(u)=e^{-1 /u }$. Here, $F(u, \bar{x})$ represents a smooth bounded function within the interval $u \in [0,1]$ subject to the constraint $\frac{\partial F(u, \bar{x})}{\partial u} < \frac{1}{u^2} + \frac{1}{(1-u)^2}$. These requirements arise from the definition of a smooth function. In this case, the associated smearing function is $\mathcal{H}_b = S(u_L+1, \bar{x}) - S(u_R, \bar{x})$.  This choice of $\mathcal{H}_b$ explicitly satisfies the conditions detailed below \eqref{difference action}, rendering actions \eqref{action begin1} and \eqref{extended action} equivalent up to $\mathcal{O}(\epsilon)$, while allowing $\mathcal{H}_b$ to incorporate an undetermined $\bar{x}$-dependent function $F$. As illustrated in the right panel of FIG. \ref{fig1}, we draw the contour of $\mathcal{H}_b$ in the example of \eqref{example_G} by setting $F=0$.

Moreover, we can obtain the following expansions for derivatives of $G_{L,R}$:
\begin{subequations}\label{main expansion one}
    \begin{align}
        &\frac{{\rm d}\mathcal{H}_b}{{\rm d} b_L}=-\frac{1}{\epsilon}\frac{{\rm d} G_L}{{\rm d}u_L}=-{\cal O}_T^L \Delta(b-b_L)=-{\cal H}_b(b_L,\epsilon){\cal O}_T^L\Delta(\mathcal{B}-\mathcal{B}_L),\label{left expansion}\\
        &\frac{{\rm d}\mathcal{H}_b}{{\rm d}b_{R}}=\frac{1}{\epsilon}\frac{{\rm d} G_R}{{\rm d}u_R}= {\cal O}_T^R \Delta(b-b_R)={\cal H}_b(b_R,\epsilon) {\cal O}_T ^R\Delta({\cal B}-{\cal B}_R).\label{right expansion}
    \end{align}
\end{subequations}
Here, we introduce the $T$-type operators defined as
\begin{align}\label{OT operator}
    {\cal O}_T^L:=\sum_{n=0}^\infty\frac{M_n^L}{n!}\partial_b^n,\quad  {\cal O}_T^R:=\sum_{n=0}^\infty\frac{ M_n^R}{n!}\partial_b^n,
\end{align}
where $M_n^{L,R}:=(-1)^n\epsilon^n\int_{-\infty}^\infty u_{L,R}^n\frac{{\rm d}G_{L,R}(u_{L,R},\bar x)}{{\rm d}u_{L,R}}{\rm d}u_{L,R}$. To derive the final equalities in \eqref{left expansion} and \eqref{right expansion}, the singularity of the Dirac delta function is shifted from $b_{L,R}$ to $\mathcal{B}_{L,R}$ using the property $\Delta(f(x))=\frac{\Delta(x-x_0)}{|f^{\prime}(x_0)|}$. To express the derivative $\frac{\partial{\cal H}_b}{\partial \epsilon}$, we also require $S$-type operators ${\cal O}_S^L$ and ${{\mathcal O}}_S^R$, whose explicit expressions are
\begin{align}\label{OS operator}
\mathcal{O}_S^L:=\sum_{n=0}^\infty\frac{{W}_n^L}{n!}\partial_b^n,\quad     {{\mathcal O}}^R_S:=\sum_{n=0}^\infty\frac{{W}_n^R}{n!}\partial_b^n,
\end{align}
with ${W}^{L,R}_n:=(-1)^n\epsilon^n\int_{-\infty}^\infty u_{L,R}^{n+1}\frac{{\rm d}G_{L,R}(u_{L,R},\bar x)}{{\rm d}u_{L,R}}{\rm d}u_{L,R}$. Then we obtain
\begin{align}\label{main expansion two}
    \frac{{\rm d}\mathcal{H}_b}{{\rm d}\epsilon}=-\mathcal{O}_S^L\Delta(b-b_L)+\mathcal{O}_S^R\Delta(b-b_R).
\end{align}
All coefficients $M_n^{L,R}$ and $W_n^{L,R}$ are field-dependent functions of $\bar {x}$ because $\epsilon$ is assumed to be field-dependent.

Combining the above ${\cal O}_T^{L,R}$ and ${\cal O}_S^{L,R}$ with the relation $(x-x_0)\partial_x^n\Delta(x-x_0)=-n\partial_x^{n-1}\Delta(x-x_0)$, we derive the relationship between the two types of operators as
\begin{align}\label{LS operator relation}
    (b-b_{L,R})\mathcal{O}_T^{L,R}\Delta(b-b_{L,R})=\epsilon\mathcal{O}_S^{L,R}\Delta(b-b_{L,R}),
\end{align}
which, together with \eqref{main expansion two}, are further proven in Appendix \ref{app:Calculation of Expansion Coefficients}.  Furthermore, the derivative $\frac{{\rm d} {\cal H}_b}{{\rm d} \epsilon}$ can be rewritten as a function of the derivatives with respect to $b_L$ and $b_R$:
\begin{align}\label{derivative to e}
  \begin{gathered}
\begin{gathered}
\frac{{\rm d}\mathcal{H}_{b}}{{\rm d}\epsilon}=-\mathcal{O}_S^L\Delta(b-b_L)+\mathcal{O}_S^R\Delta(b-b_R).
\end{gathered}
\end{gathered}
\end{align}

To further identify the MCFs at the boundaries, we first introduce the field-space one forms
\begin{align}
    \chi_L:=\delta b_L+\frac{b-b_L}{\epsilon}\delta\epsilon,\quad \chi_R:=\delta b_R+\frac{b-b_R}{\epsilon}\delta\epsilon,
\end{align}
which satisfy
\begin{subequations}\label{new chi relation}
\begin{align}
    & \mathcal{O}_T^L\delta b_L+\mathcal{O}_S^L\delta\epsilon=\chi_L\mathcal{O}_T^L\Delta(b-b_L),\\
    & \mathcal{O}_T^R\delta b_R+\mathcal{O}_S^R\delta\epsilon=\chi_R\mathcal{O}_T^R\Delta(b-b_R).
\end{align}
\end{subequations}
 Near the left and right boundaries, $\delta \mathcal{H}_b$ can be expressed in terms of $\chi_L$ and $\chi_R$ as
\begin{align}\label{deltaHchi}
\begin{aligned}
{\left. {\delta {{\cal H}_b}} \right|_{b = {b_{L}}}} =-\frac{1}{\epsilon}\chi_L\frac{{\rm d}G_{L}}{{\rm d}u_{L}},~~~~~{\left. {\delta {{\cal H}_b}} \right|_{b = {b_{R}}}} =\frac{1}{\epsilon}\chi_R\frac{{\rm d}G_{R}}{{\rm d}u_{R}}.
\end{aligned}
\end{align}
 which do not hold for $\chi$ because $\mathcal{H}_b$ is not only a function of ${\cal B}$. The variation of $\delta \mathcal{B}$ at the boundaries can be prescribed as being driven by the fluctuating boundary conditions:
 \begin{align}\label{new boundary con}
     \delta {\cal B}_{L,R}=-\chi^b_{L,R}\partial_b {\cal B}_{L,R},
 \end{align}
which implies that only at the leading order of $b-b_{L,R}$ are the DRF fluctuations captured by $\delta b_{L,R}$. At linear order, these fluctuations correspond to $\delta \epsilon$, a contribution that cannot be ignored once the operator corrections from \eqref{new chi relation} are included. We denote the coordinates of $x^\mu$ excluding $\cal B$ and $\cal C$ by $\overline{x}^{\bar i}$, and consider that  $y^A$ is field-independent, given by $\delta\bar x^{\bar i}=0$. Considering  the associated map $ U(\bar x^{\bar i})=\bar X^{\bar I}=\bar x^{\bar i}$\footnote{It is worth noting that $U$ can be defined via alternative mappings, potentially yielding richer boundary dynamics, analogous to the boundary diffeomorphisms considered in \cite{Joung:2023doq}.} and taking $ U(\bar x)$ to be the identity map without loss of generality, we then obtain the coordinates $ x^\mu$ and $y^A$ in \eqref{MCF in component} as
\begin{align}\label{coordinate componets}
	 x^\mu=(b,c,\overline{x}^{\bar i}),\quad  y^A=U^A=({\cal B},{\cal C},\bar X^{\bar I}).
\end{align}
To compute the MCFs, we first calculate $\delta \mathcal{B}$. Using the integral decomposition $\int_{b_0}^b=\int_{b_0}^{b_{L,R}}+\int_{b_{L,R}}^b$, we find that near the two boundaries, $\delta \mathcal{B}$ satisfies
\begin{align}\label{dB}
    \delta\mathcal{B}=\delta\mathcal{B}_{L,R}+\delta\left(\int_{b_{L,R}}^b\mathcal{H}_b(\tau)\mathrm{d}\tau\right).
\end{align}
Using this expression, the MCF components in \eqref{MCF in component}, and the boundary conditions \eqref{new boundary con}, we find that $\chi^{\overline{x}} = 0$, while the remaining non-zero component is given by
\begin{align}\label{MC constarint}
    \lim_{b\to b_{L,R}}\chi^b=-\frac{\delta{\cal B}_{L,R}+\int_{b_{L,R}}^b\delta\mathcal{H}_b(\tau)\mathrm{d}\tau}{{\cal H}_b}=-\frac{\delta{\cal B}_{L,R}}{{\cal H}_b}-\lim_{b\to b_{L,R}}\frac{\delta\mathcal{H}_b(b)}{\frac{{\rm d}\mathcal{H}_b}{{\rm d}b}}\frac{{\cal H}_b-1}{{\cal H}_b}\approx\chi_{L,R},
\end{align}
where the second equality is obtained by applying L'Hôpital's rule, combining $\left. {\frac{{\rm d}\mathcal{H}_b}{{\rm d}b}} \right|_{b= b_L}=\frac{1}{\epsilon}\frac{{\rm d}G_{L}}{{\rm d}u_{L}}$ and $\left. {\frac{{\rm d}\mathcal{H}_b}{{\rm d}b}} \right|_{b= b_R}=-\frac{1}{\epsilon}\frac{{\rm d}G_{R}}{{\rm d}u_{R}}$ with the fact that both $\int_{b_{L,R}}^b\delta\mathcal{H}_b(\tau)\mathrm{d}\tau$ and $\delta{\cal H}_b$ are small quantities at the boundaries. By virtue of this equivalence with the MCF subject to specific boundary conditions, $\chi_{L,R}$ are termed \textit{linear MCFs}.

Our DRF approach differs from existing standard approaches in two main ways. First, in \eqref{dH}, we decompose the dynamical reference frame into three fundamental fields generating boundary dynamics. Similar to \cite{Goeller:2022rsx}, we employ smearing functions to guarantee gauge invariance for local observables; however, we introduce $b_{L,R}$ and $\epsilon$ as fundamental fields via an action extension to capture soft-cutoff boundary fluctuations rather than using bulk matter fields. Second, we utilize T-type and S-type operators to implement soft cutoffs in covariant theories, rather than using bulk operators to enforce micro-causality \cite{Donnelly:2015hta,Donnelly:2016rvo,Hoehn:2025pmx}. These operators manifest in the derivatives of $\mathcal{H}_b$ and preserve diffeomorphism invariance, owing to the field dependence of $\epsilon$. These aspects will be detailed in the subsequent analysis of symplectic potentials and charges. Finally, our thickness field parallels the split structure in \cite{Donnelly:2017jcd,Giddings:2019hjc}, which bypasses Hilbert space non-factorizability by defining subsystems as distinguishable only within a finite region. A similar conclusion emerges: actions at the thickness field scale are indistinguishable via \eqref{difference action}; thus, the $\epsilon$ field effectively sets both the perturbation scale and the resolution limit for the theory.

\section{Covariant phase space formalism}
\label{sec:Covariant phase space with the fluctuating and smearing boundary}
\subsection{Symplectic potentials}
\label{subsec:symplectic potentials and symplectic forms}

We now perform the variation of the action for the covariant theory, subject to the boundary conditions proposed below \eqref{difference action}. In this derivation, we treat $U_c$ as the identity map on $c$, while $U_b$ and $\mathcal{H}_b$ take the forms given in \eqref{U_distri} and \eqref{smearing function}, subject to the boundary conditions in \eqref{new boundary con}. Let us first consider the case in the absence of the boundary Lagrangian $\ell$. According to \eqref{extended action}, we have
\begin{align}
	\begin{aligned}\label{variation of action}
		\delta {S }_\epsilon = \int_{\cal M} {{{\rm{d}}^D} x} \left( \delta {\cal L}{{\cal H}_b} + {\cal L}\delta{\cal H}_b \right)
		 \simeq \int_{\cal M} {{{\rm{d}}^D}x} \left( -\Theta^\mu_{LW}\partial_\mu{{\cal H}_b} +{\cal L}\delta{\cal H}_b\right).
	\end{aligned}
\end{align}
 The symbol ``$\simeq$'' denotes on-shell equality, i.e., equality holds when the equation of motion $E_\Phi=0$ is satisfied. This implies that although the theory is expanded by an order of $\epsilon$ in the form of \eqref{difference action}, the constraints in the bulk remain unchanged. We then further express $\partial_{\mu}{\cal H}_b$ in terms of the following normal vectors:
\begin{subequations}
\begin{align}
&N_\mu^L:=-\left(\delta_\mu^b-\partial_\mu b_L\right),\\
&N_\mu^R:=\delta_\mu^b-\partial_\mu b_R,
\end{align}
\end{subequations}
which arise from the differentiation of the boundary scalar fields $\phi_{L}:=-(b-b_{L})$ and $\phi_{R}:=b-b_{R}$.
Then using the derivative of $\mathcal{H}_b$ with respect to $b_{L,R}$ in \eqref{main expansion one},  the derivative $\partial_\mu\mathcal{H}_b$ can be expressed as
\begin{align}\label{xmu of H}
\begin{aligned}
\partial_\mu{\mathcal{H}_b}&=\frac{{\partial}G_L}{{\partial}u_L}\frac{\partial u_L}{\partial x^\mu}-\frac{{\partial}G_R}{{\partial}u_R}\frac{\partial u_R}{\partial x^\mu}\\
&=\frac{{\rm d}\mathcal{H}_b}{{\rm d} b_L}N_\mu^L-\frac{{\rm d}\mathcal{H}_b}{{\rm d} b_R}N_\mu^R+\frac{{\rm d}\mathcal{H}_b}{{\rm d} \epsilon}\partial_\mu\epsilon.
\end{aligned}
\end{align}
 For the variation $\delta{\cal H}_b$, according to \eqref{delta basis}, \eqref{main expansion one} and \eqref{derivative to e}, we can rewrite the expressions in \eqref{variation of action} using the T-type and S-type operators as
\begin{subequations}
    \begin{align}
        -\Theta_{LW}^\mu\partial_\mu {\cal H}_b&=\Theta_{LW}^\mu( N_\mu^L{\cal O}_T^L+\partial_\mu\epsilon{\cal O}_S^L)\Delta(b-b_L)+\Theta_{LW}^\mu( N_\mu^R{\cal O}_T^R-\partial_\mu\epsilon{\cal O}_S^R)\Delta(b-b_R),\\
        \delta {\cal H}_b&=-\left({\cal O}_T^L\delta b_L+{\cal O}_S^L\delta\epsilon\right)\Delta(b-b_L)+\left({\cal O}_T^R\delta b_R+{\cal O}_S^R\delta\epsilon\right)\Delta(b-b_R).
    \end{align}
\end{subequations}
Subsequently, by using \eqref{LS operator relation} to convert the $S$-type operator into the $T$-type operator, we can rewrite \eqref{variation of action} as
\begin{align}\label{dS}
\delta S_{\epsilon} \simeq    \int_{{\cal B}_L} {\rm d}^{D-1} \bar{x} \overline{\mathcal{O}}_T^L\left(\Theta_{L W}^\mu {N}_\mu^L-\chi_L\mathcal{L} \right)  +\int_{{\cal B}_R} {\rm d}^{D-1} \bar{x} \overline{\mathcal{O}}_T^R\left(\Theta_{L W}^\mu {N}_\mu^R+\chi_R\mathcal{L} \right).
\end{align}
Here, we impose the condition $\mathcal{H}_b \approx 1$ in the interval $[b_{L},b_R]$ and shift the singularity of the Dirac delta function from the spacetime boundaries $b_{L,R}$ to the relational boundaries $\mathcal{B}_{L,R}$. This implies that we are studying covariant subsystems.

To further simplify expression \eqref{dS}, we first
denote the integrals over surfaces ${\cal B}={\cal B}_{L,R}$ by $\int_{{{\cal B}_{L,R}}} {} $. Then we define the new operators
    \begin{align}
    \begin{aligned}
\overline{\mathcal{O}}_T^L:=\sum_{n=0}^\infty\frac{(-1)^nM_n^L}{n!}\partial_b^n,\quad\overline{\mathcal{O}}_T^R:=\sum_{n=0}^\infty\frac{(-1)^nM_n^R}{n!}\partial_b^n,
\end{aligned}
    \end{align}
and
\begin{align}\label{new sym}
&\widehat{\Theta}_{L,R}^\mu:=\Theta_{LW}^\mu+\chi_{L,R}^\mu\mathcal{L},
\end{align}
where the linear DRFs $\chi_L^\mu$ and $\chi_R^\mu$ satisfy the relations $\chi_L=-\chi_L^\mu N_\mu^L$ and $\chi_{R}=\chi_{R}^{\mu}N_{\mu}^{R}$. Subsequently, using \eqref{U_distri} and \eqref{xmu of H}, the derivative of $\mathcal{B}$ is given by $\partial_{\mu}\mathcal{B}=\mathcal{H}_{b}(b,x)\delta_{\mu}^{b}+\int_{b_{0}}^{b}\partial_{\mu}\mathcal{H}_{b}(\tau,x)\mathrm{d}\tau$ such that  we obtain
\begin{subequations}\label{partal B}
    \begin{align}
        & {\left. {\partial_{\mu}{\cal B}} \right|_{b = {b_{L}}}}=-\left(\delta_\mu^b-\partial_\mu b_{L}\right)=- N_\mu^{L},\\
        &{\left. {\partial_{\mu}{\cal B}} \right|_{b = {b_{R}}}}=\delta_\mu^b-\partial_\mu b_{R}= N_\mu^{R},
        \end{align}
\end{subequations}
in which we have considered that the integrals $\int^b_{b_{L,R}}{\cal H}_b{\rm d}\tau$ vanish at $b=b_{L,R}$ due to the boundary conditions of ${\cal H}_b$. These results mean $\partial_\mu\mathcal{B}$ serves as the normal vectors at the boundaries. Finally, plugging \eqref{new sym} and \eqref{partal B} into \eqref{dS} gives us a concise form for $\delta S_\epsilon$:
\begin{align}\label{left right symplectic potential}
    \delta S_\epsilon\simeq-\int_{{\cal B}_{L}}{\rm d}^{D-1}{x}\overline{\mathcal{O}}_{T}^{L}\left(\widehat{\Theta}_{L}^{\mu}\partial_{\mu}\mathcal{B}\right)+\int_{{\cal B}_{R}}{\rm d}^{D-1}{x}\overline{\mathcal{O}}_{T}^{R}\left(\widehat{\Theta}_{R}^{\mu}\partial_{\mu}\mathcal{B}\right),
\end{align}
where, due to the condition $\mathcal{H}_b \approx 1$ in the interval $[b_L,b_R]$, the Dirac delta functions $\Delta(\mathcal{B})$ and $\Delta(b)$ become interchangeable.

Similar steps can be employed to calculate the corrections introduced by the ambiguity term. Because of $-\delta(\partial_\mu {\cal H}_b\ell^\mu)=-\partial_\mu \delta{\cal H}_b\ell^\mu-\partial_\mu {\cal H}_b\delta\ell^\mu$, the part of the symplectic potential associated with $\ell$ is
\begin{align}\label{ambi sym}
   \delta S^{(\ell)}_\epsilon=-\int_{{\cal B}_L}{\rm d}^{D-1} x\overline{\cal O}^L_T\Theta^L_\ell+\int_{{\cal B}_L}{\rm d}^{D-1} x\overline{\cal O}^R_T\Theta^L_\ell,
\end{align}
where
\begin{align}
        &\Theta_{\ell}^{L,R}:=N_{\mu}^{L,R}(\delta\ell^{\mu}+\chi_{L,R}^\mu\partial_{\nu}\ell^{\nu}).
    \end{align}
Consequently, it is obvious that the leading-order form of the integral over ${\cal B}_{L,R}$ in $\delta S_\epsilon+\delta S^{(\ell)}_\epsilon$ is consistent with the symplectic potential pulsing the ambiguities of boundary Lagrangian in \cite{Adami:2024gdx}.

Overall, in this subsection, relaxing the boundary conditions in equation \eqref{relational boundary condition} permits the boundary fluctuations introduced by equation \eqref{new boundary con}, invalidating the equivalence between $\chi$ and $\chi_{L,R}$ in \eqref{MC constarint}. Yet, according to \eqref{left right symplectic potential}, $\chi_{L,R}$ and the operators $\overline{\mathcal{O}}_T^{L,R}$ still combine within the symplectic potentials. Consequently, because we incorporate the linear MCFs into the symplectic potentials rather than the MCFs themselves, the resulting Lagrangian variation differs from \eqref{extended symplectic potential}. Through the DRF, we can identify the divergence in the symplectic potential controlled by the coefficients $M_n^{L,R}$ in \eqref{OT operator}. This enables us to avoid possible divergent behavior of the differential forms at the spacetime boundaries. We will see more specific examples in Section \ref{sec:Noether charge at the asymptotic boundary of general relativity in the FG gauge}.

\subsection{Symplectic forms and integrable charges}
\label{subsec:Symplectic forms and integrable charges}
Let us first consider the symplectic form without the ambiguity term. Based on the normal vector relations in \eqref{partal B}, we introduce the left boundary corner as a compact region defined by $\mathcal{S}_L:=\{x^\mu|\:\mathcal{C}=0,\mathcal{B}=0\}=\mathcal{B}\cap\mathcal{C}$, along with the binormal $\mathcal{E}_{\mu\nu}:=2\partial_{[\mu}\mathcal{C}\partial_{\nu]}\mathcal{B}$ defined on it.

To construct the symplectic form, we establish a variational rule starting with covariant scalars $P_\epsilon(\mathcal{B})$ and $N_\epsilon(\mathcal{C})$. These satisfy $\delta_{\chi_L} P_\epsilon=0$ and $\delta_{\chi_L} N_\epsilon=0$, which, due to \eqref{covariant variantion of B}, are equivalent to ${\chi}_L^{\mu}\partial_{\mu}P_\epsilon =-\delta P_\epsilon$ and ${\chi}_L^{\mu}\partial_{\mu}N_\epsilon =-\delta N_\epsilon$. Then the variation we calculate takes the form
\begin{align}\label{general delta X}
	{\cal X}=\delta\left(\int_{\cal M}{\rm d}^DxXP_\epsilon({{\cal B}})N_\epsilon({{\cal C}})\right),
\end{align}
where $X$ is a one-form in field space satisfying ${\cal L}_{\chi }X=\partial_\mu (\chi_L^\mu\wedge{ X})$ (examples include $\widehat{\Theta}_L^\mu$ and $\widehat{\Theta}_R^\mu$ on co-dimension one surfaces)\footnote{This formalism relies on the fact that $X(x){\rm d}^{\bar D}x=X(x){\rm d}x^1\wedge \cdots \wedge {\rm d}x^{\bar D}$ is a top form on the $\bar D$-dimensional space. Applying Cartan's magic formula, we obtain $\mathcal{L}_{\chi_L}({X(x){\rm d}^{\bar D}x})=({\rm d}i_{\chi_L}+i_{\chi_L} {\rm d}){X(x){\rm d}^{\bar D}x}={\rm d}i_{\chi_L}({X(x){\rm d}^{\bar D}x})=\partial_\mu(\chi_L^\mu\wedge X){\rm d}^Dx$, where we used the property that the exterior derivative of a top form vanishes. Furthermore, since both $\chi_L^\mu$ and $X$ are one-forms in field space, their product implies a wedge product.}.   Ignoring the total derivative terms, we obtain the following result using integration by parts
\begin{align}\label{variationrule}
\begin{aligned}
\mathcal{X}&=\int_{\cal M}{\rm d }^{D}{x}P_{\epsilon}N_{\epsilon}\delta_{\chi_L}X.
\end{aligned}
\end{align}
A direct application of this variational rule is realized when we set $P_\epsilon=\Delta(\mathcal{B})$, $N_\epsilon=\Delta(\mathcal{C})$, and $X=Z^{\mu\nu}\partial_\mu\mathcal{B}\partial_\nu\mathcal{C}$, where $Z^{\mu\nu}$ is explicitly identified as an antisymmetric tensor that simultaneously acts as a one-form in field space. This yields the relation
\begin{align}\label{z variation}
   \delta\int_{\mathcal{S}_L}\mathrm{d}^{D-2}x {\cal E}_{\mu\nu}Z^{\mu\nu}&=\int {\rm d}^D x\Delta({\cal B})\Delta({\cal C})[\delta(Z^{\mu\nu}\partial_\mu{\cal B}\partial_\nu{\cal C})+\partial_\alpha(\chi_L^\alpha\wedge Z^{\mu\nu}\partial_\mu{\cal B}\partial_\nu{\cal C})]\notag\\
   &=\int_{\mathcal{S}_L}\mathrm{d}^{D-2}x {\cal E}_{\mu\nu}\left(\delta Z^{\mu\nu}-2\partial_\alpha Z^{\mu\alpha}\wedge\chi_L^\nu\right),
\end{align}
where we have apply the identities $\delta ({\cal E}_{\mu\nu}{Z}^{\mu \nu})+\partial_\mu\left({\chi}_L^\mu \wedge {\cal E}_{\mu\nu}{Z}^{\mu \nu}\right)=\left(\delta {Z}^{\mu \nu}+\mathcal{L}_{{\chi}_L} {Z}^{\mu \nu}\right) {\cal E}_{\mu\nu}$ with  Lie derivative $\mathcal{L}_{\chi_{L}}Z^{\mu\nu}=-2\partial_{\alpha}Z^{[\mu|\alpha|}\wedge\chi_{L}^{\nu]}+\partial_{\alpha}(3\chi_{L}^{[\alpha}\wedge Z^{\mu\nu]})$. This relation will be used to calculate the contribution of the boundary Lagrangian $\ell$ to the symplectic form.

We now compute the symplectic form corresponding to symplectic potential \eqref{left right symplectic potential}. We consider the zeroth-order contribution in $\epsilon$ by applying \eqref{variationrule} and take the left boundary as the example, the corresponding symplectic form $\Omega$ is
\begin{align}\label{symplectic form example}
\Omega=\delta\int_{\mathcal{M}}{\rm d}^D x{(-\partial_\mu {\cal B}\widehat{\Theta}_L^\mu)}{\Delta({\cal B})}{H({\cal C})},
\end{align}
where $\widehat\Theta_L^\mu=\Theta_{LW}^\mu+\chi_L^\mu {\cal L}$. Comparing with \eqref{general delta X}, we can read off $X=-\partial_\mu{\cal B}\widehat\Theta_L^\mu$, $P_\epsilon=\Delta({\cal B})$ and $N_\epsilon=H({\cal C})$. Applying the variational rule from \eqref{variationrule} to $\Omega$, the first part arising from \eqref{symplectic form example} is
\begin{align}\label{variation A}
\begin{aligned}
	& \int_{\cal M} {\rm d}^D x\Delta({\cal B})H({\cal C})\delta (-\partial_\mu {\cal B}\widehat{\Theta}_L^\mu)\\
    =&\int_{\cal M} {\rm d}^D x{\Delta({\cal B})}H({\cal C})(-\partial_\mu\delta {\cal B}\wedge\widehat{\Theta}_L^\mu)
    +\int _{\cal M}{\rm d}^D x{\Delta({\cal B})}H({\cal C})(-\partial_\mu {\cal B}{\delta}\widehat{\Theta}_L^\mu)\\
    =&\int_{\cal M} {\rm d}^D x{\Delta({\cal B})}H({\cal C})(-\partial_\mu\delta {\cal B}\wedge{\Theta}_{LW}^\mu)
    -\int _{\cal M}{\rm d}^D x{\Delta({\cal B})}H({\cal C})\partial_\mu {\cal B}{\delta}{\Theta}_{LW}^\mu+\int _{\cal M}{\rm d}^D x{\Delta({\cal B})}H({\cal C})\delta{\cal L}\wedge\delta {\cal B}.
    \end{aligned}
\end{align}
The third equality follows from $\delta \chi_L^\mu=0$ at $b=b_L$. Recalling the relation $\chi_L^\mu \partial_\mu \bar\Phi = -\delta \bar\Phi$ (from Eq. \eqref{covariant variantion of B}) for $\bar\Phi \in \{\mathcal{B}, \mathcal{C}\}$, we find that $\chi_L^\mu \partial_\mu (\delta \bar\Phi) = -\delta \chi_L^\mu \partial_\mu \bar\Phi = 0$ (based on $\delta_{\chi_L}{\cal B}=0$ and  $\delta_{\chi_L}{\cal C}=0$). The second part of \eqref{symplectic form example} is calculated as
\begin{align}\label{variation BC}
	\begin{aligned}
&\int_{\cal M}{\rm d}^D x\Delta({\cal B}-{\cal G}_L)H({\cal C})\partial_\mu(-\chi_L^\mu\wedge\partial_\nu {\cal B}\widehat{\Theta}^\nu)\\
=&-\int_{\cal M}{\rm d}^D x[\partial_{{\cal B}}\Delta({\cal B})H({\cal C})\delta{\cal B}+\Delta({\cal B})\Delta({\cal C})\delta{\cal C}]\wedge\Theta_{LW}^\mu\partial_\mu{\cal B}+\int_{\cal M}{\rm d}^Dx\Delta({\cal B})\Delta({\cal C})\delta{\cal C}\wedge\delta{\cal B}{\cal L}.
	\end{aligned}
\end{align}
  {Combining \eqref{variation A} and \eqref{variation BC} yields the symplectic form at zeroth order in $\epsilon$}
\begin{align}\label{sym zero}
	\begin{aligned}
		\Omega&=\int_{{\cal B }_L^>} {\rm d}_{\mu}\bar x\delta\Theta_{LW}^\mu+\int _{\cal M}{\rm d}^D x{\Delta({\cal B})}\Delta({\cal C})Y_\mu\wedge{\Theta}_{LW}^\mu+\int_{\cal M}{\rm d}^D x\Delta({\cal B})\Delta({\cal C})\delta{\cal C}\wedge\delta{\cal B}{\cal L},\\
	\end{aligned}
\end{align}
where we set ${\rm d} \bar x_\mu=-\partial_\mu{\cal B}{\rm d}^{D-1}\bar x$, $Y_\mu=\partial_\mu{\cal C}\delta{\cal B}-\partial_\mu{\cal B}\delta{\cal C}$ and $\mathcal{B}_{L}^>$ denotes the intersection of these surfaces with the region defined within the interval ${\cal C}\ge0$.

For the full orders of the symplectic form in $\epsilon$, we have
\begin{align}\label{symplectic form}
\begin{aligned}
\Omega_\epsilon=\delta\int_{\mathcal{M}}{{\rm d}^{D-1}\bar x}\Delta({{\cal B}})H({\mathcal{C}})\overline{\mathcal{O}}\widehat\Theta_L,
\end{aligned}
\end{align}
where $\widehat\Theta_L=-\widehat\Theta^\mu_L\partial_\mu{{\cal B}}$ and $\overline{\mathcal{O}}$ denotes $\overline{\mathcal{O}}_T^L$ for simplicity. Similar rules can be applied to the right boundary. Then, following steps similar to those used to obtain $\Omega$, taking into account the field dependence of the operator $\overline{\mathcal{O}}$, and reapplying the variational rule \eqref{variationrule}, $\Omega_\epsilon$ can be further calculated as
\begin{equation}\label{main symplectic form}
\begin{aligned}
\Omega_\epsilon=\Omega_{(1)}+\Omega_{(2)}+\Omega_{(3)},
\end{aligned}
\end{equation}
in which we have
\begin{subequations}
    \begin{align}
        \Omega_{(1)}&=-\int_\mathcal{M} {\rm d}^Dx\, \Delta(\mathcal B)H(\mathcal C)\,\overline{\mathcal O}\,
\bigl(\delta\Theta_{LW}^\mu\partial_\mu\mathcal B\bigr)+\int_\mathcal{M} {\rm d}^Dx\, \Delta(\mathcal B)\Delta(\mathcal C)\,\overline{\mathcal O}\,
\bigl(Y_\mu\wedge\Theta_{LW}^\mu+\delta{\cal C}\wedge\delta{\cal B}{\cal L}\bigr),\label{O1}\\
\Omega_{(2)}&= -\int_\mathcal{M} {\rm d}^Dx\, \Delta({\cal B}) H({\cal C})\,(\partial_\mu\overline{\mathcal O})
(\widehat\Theta_{L}^\mu)\wedge\delta\mathcal B,\label{O2}\\
\Omega_{(3)}&=\int_{\mathcal B_L^>} {\rm d}^{D-1}\bar x\,
\delta\overline{\mathcal O}\wedge\widehat{\Theta}_L,\label{O3}
\end{align}
\end{subequations}
where we consider that, although $\partial_\mu$ and $\overline{\mathcal{O}}$ do not commute, $\partial_\mu{\cal B}$ commutes with $\overline{\mathcal{O}}$ according to \eqref{partal B}. Comparing \eqref{sym zero} and \eqref{main symplectic form}, the leading terms of $\Omega_{(1)}$ recover the original symplectic form $\Omega$ \cite{Adami:2024gdx,Golshani:2024fry}. Meanwhile, $\Omega_{(2)}$ and $\Omega_{(3)}$ stem from non-commutativity of $\overline{\mathcal{O}}$ with partial derivatives and its field dependence, respectively.

We now contract the full-order symplectic form $\Omega_\epsilon$ along the $\xi$ direction in field space by using the contraction $\iota_\xi$. To this end, we first construct the displacement relation associated with the T-type operators. To express the operator $\overline{\mathcal{O}}$ in the exponential form $\overline{\mathcal{O}} = e^{\mathcal{K}}$, we introduce $\kappa_n^{L} := \frac{\mathrm{d}^n}{\mathrm{d}b^n}\left[\ln \overline{\mathcal{O}}\right]|_{b=0}$. Here, the generating function is defined as $\overline{\mathcal{O}} := \sum_{n=0}^\infty \frac{M_n^{L}}{n!} (-\partial_b)^n$, and the coefficients $M_n^{L}$ are given below \eqref{OT operator}. By defining the exponent as $\mathcal{K} := \sum_{n=0}^\infty \frac{(-1)^n \kappa_n}{n!} \partial_b^n$, this formulation ensures that the relationship between $M_n^L$ and $\kappa_n$ is uniquely determined. Specifically, the first two orders of $\kappa_n$ satisfy $\kappa_1=M_1^L$ and $\kappa_2=M_2^L - {(M_1^L)^2}$. This allows us to interpret the operator as actively displacing the $\text N$-form field $\mathcal{F}$ within the field space. Specifically, the action of shifting $\mathcal{F}$ in this manner satisfies the following displacement relation:
\begin{align}\label{shifted relation}
\overline{\cal O}{\cal F}({b}) = {\cal F}(b_{\cal F}),
\end{align}
where $b_{\cal F}$ represents the shifted location at the left boundary\footnote{The field-space one-form $\mathcal{F}$ can legitimately appear in the denominator because, following the field expansion in \eqref{delta basis}, its dependence on $b$ at the left boundary decouples from the field-space variations. This separability allows us to express a scalar function and a purely geometric form, ${\cal F}=\bar f(b,\bar x)\delta b_{\cal F}$, ensuring that the field-dependent factors in the numerator and denominator of $\frac{\partial_b^k{\cal F}}{\partial_b{\cal F}}$ perfectly cancel each other out.}
\begin{align}\label{bG solution}
    b_{\cal F}:= b_{L}+\sum_{k=1}^\infty\frac{(-1)^k}{k!}\kappa_k\left.\frac{\partial_b^k{\cal F}}{\partial_b{\cal F}}\right|_{b=b_{L}}.
\end{align}
Applying transformation $\delta_\xi$ derivative $\partial_\mu$ to $\overline{\cal O}{\cal F}({b})$ and combining it with the chain rules
\begin{subequations}
\begin{align}
\partial_{\mu}(\overline{\cal O}\mathcal{F})&=\partial_\mu\mathcal{F}({b_{{\cal F}}})+\partial_b\mathcal{F}(b_{\cal F})\partial_\mu b_{{\cal F}},\\
        \delta\left[\overline{\cal O}\mathcal{F}\right]&={\delta\mathcal{F}}(b_{\cal F})+(-1)^{\text N}\left[\partial_b\mathcal{F}\right](b_{\cal F})\delta b_{\cal F},
    \end{align}
    \end{subequations}
we can further derive
\begin{subequations}\label{trans_of_operator_mai}
    \begin{align}
    &\partial_{\mu}(\overline{\mathcal{O}}\mathcal{F})=[\mathcal{F}](b_{\cal F})\partial_{\mu}b_{\cal F},\\
&(\delta\overline{\mathcal{O}})\mathcal{F} = (-1)^{\text N}\left[\partial_b\mathcal{F}\right](b_{\cal F})\wedge \delta b_{\cal F}.
    \end{align}
    \end{subequations}
Particular attention should be drawn to the second result above. This stems entirely from the fact that the field dependence of $\cal F$ arises exclusively from $b_{\mathcal{F}}$, a property we refer to as \textit{pointwise dependence}.

Besides establishing these relations, we also need the definition of the Noether current $J_\xi^\mu:=\Theta _{LW}^\mu(\delta_\xi)-\xi^\mu{\cal L}$. We identify the Noether charge $Q_N(\xi) = \frac{1}{2} {\cal E}_{\mu\nu} Q^{\mu\nu}_N(\xi)$, where $Q^{\mu\nu}_N(\xi)$ is an antisymmetric tensor and satisfies $J_\xi^\mu=\partial_{\nu}Q_N^{\mu\nu}(\xi)$. Moreover, combining the field transformations $\delta_\xi\mathcal{B}=\xi^\mu\partial_\mu\mathcal{B}$ and $\delta_\xi\mathcal{C}=\xi^\mu\partial_\mu\mathcal{C}$ with the MCF definition in \eqref{MCF in component} directly yields $(\iota_\xi\chi_L^\mu)\partial_\mu\mathcal{B}=-\xi^\mu\partial_\mu\mathcal{B}$.
Subsequently,  $-\iota_\xi\Omega_{(1)}$ can be evaluated as
\begin{align}\label{contrac 1}
-\iota_{\xi}\Omega_{(1)} = & \delta Q_{c} -\delta_{\overline{\mathcal{O}}}{Q_{c}} +\delta_{{Q_N}}{Q_{b}} -\int_{\mathcal{S}_L}{{\rm d}^Dx (\partial_\nu\overline{\mathcal{O}}) \big( Q_N^{\mu\nu}(\xi) \big) }Y_\mu \notag\\
&+\int_{\mathcal{B}_L^>}\mathrm{d}^Dx\left[-(\delta_\xi\overline{\mathcal{O}})(\widehat{\Theta}_L^\mu)\partial_\mu\mathcal{B}+(\partial_\mu\overline{\mathcal{O}})(\widehat{\Theta}_L^\mu)\delta_\xi\mathcal{B}+(\delta_\xi\overline{\mathcal{O}})(\chi_L^\mu\mathcal{L})\partial_\mu\mathcal{B}-(\partial_\mu\overline{\mathcal{O}})(\chi_L^\mu\mathcal{L})\delta_\xi\mathcal{B}\right],
\end{align}
where the Noether charge density $Q_N^{\mu\nu}(\xi)$ emerges by utilizing the identities $\iota_\xi \widehat{\Theta}_L^\mu=\partial_\nu Q_N^{\mu\nu}(\xi)$, $\iota_{\xi}Y_{\mu}=\xi^{\nu}\mathcal{E}_{\mu\nu}$ and $\delta\mathcal{E}_{\mu\nu}=2\partial_{[\nu}Y_{\mu]}$. The definitions of $Q_c$ and $Q_{b}$ are given by
\begin{subequations}
\begin{align}
    &{Q}_{c}:=-\int_{\mathcal{S}_L}\mathrm{d}^Dx\overline{\mathcal{O}}(Q_N^{\mu\nu}(\xi))\partial_\mu\mathcal{C}\partial_\nu\mathcal{B},\\
    &{Q}_{b}:=-\int_{\mathcal{B}_L^>}\mathrm{d}^{D}x(\partial_{\mu}\overline{\mathcal{O}})\partial_{\nu}\mathcal{B}Q_{N}^{\mu\nu}(\xi).
\end{align}
\end{subequations}
 The operators $\delta_{{Q_N}}$ and $\delta_{\overline{\mathcal{O}}}$ denote variations taken exclusively with respect to $Q_N^{\mu\nu}(\xi)$ and $\overline{\mathcal{O}}$. Subsequently, the remaining $\epsilon$-dependent terms in $\Omega_\epsilon$ are evaluated as
\begin{subequations}\label{O2O3}
    \begin{align}
-\iota_{\xi}\Omega_{(2)}=&\int_{\mathcal{B}_l^>}\mathrm{d}^{D}x(\partial_{\mu}\overline{\cal O})(\partial_{\nu}Q_{N}^{\mu\nu}(\xi))\delta\mathcal{B}-\int_{\mathcal{B}_L^>}\mathrm{d}^{D}x(\partial_{\mu}\overline{\cal O})(\delta_{\xi}\mathcal{B}\widehat{\Theta}_{L}^{\mu})\notag\\
=&\int_{\mathcal{S}_L}\mathrm{d}^{D-1}x(\partial_\nu\overline{\mathcal{O}})(Q_N^{\mu\nu}(\xi))\partial_\mu {\cal B}\delta\mathcal{B}+\int_{\mathcal{M}}\mathrm{d}^Dx\Delta^{\prime}(\mathcal{B})H(\mathcal{C})(\partial_\nu\overline{\mathcal{O}})(Q_N^{\mu\nu}(\xi))\partial_\mu\mathcal{B}\delta\mathcal{B}\notag\\
&+\int_{\mathcal{B}_{L}^{>}}\mathrm{d}^{D}x(\partial_{\nu}\overline{\mathcal{O}})\left(Q_{N}^{\mu\nu}(\xi)\right)\partial_{\mu}(\delta\mathcal{B})-\int_{\mathcal{B}_L^>}\mathrm{d}^{D}x(\partial_{\mu}\overline{\cal O})(\delta_{\xi}\mathcal{B}\widehat{\Theta}_{L}^{\mu}),\label{O2}\\
-\iota_{\xi}\Omega_{(3)}=& \int_\mathcal{{\cal B}_L^>}\mathrm{d}^Dx(\delta_\xi\overline{\mathcal{O}})(\widehat{\Theta}_L^\mu)\partial_\mu\mathcal{B}-\int_{{\cal B}_L^>}\mathrm{d}^Dx\delta\overline{\mathcal{O}}(\partial_\nu Q_N^{\mu\nu}(\xi))\partial_\mu\mathcal{B}\notag\\
=&\int_\mathcal{{\cal B}_L^>}\mathrm{d}^Dx(\delta_\xi\overline{\mathcal{O}})(\widehat{\Theta}_L^\mu)\partial_\mu\mathcal{B}+\int_{\mathcal{B}_{L}^{>}}\mathrm{d}^{D}x(\partial_{\nu}\delta\overline{\mathcal{O}})Q_{N}^{\mu\nu}(\xi)\partial_{\mu}\mathcal{B}-\int_{\mathcal{S}_{L}}\mathrm{d}^{D-1}x(\delta\overline{\cal O})(\partial_{\mu}\mathcal{C}\partial_{\nu}\mathcal{B}Q_{N}^{\mu\nu}(\xi))\label{Omega3}.
        \end{align}
\end{subequations}
 We can first address some of the non-integrable terms on ${\cal B}_L^>$ in \eqref{contrac 1} and \eqref{O2O3}. Summing the final line of \eqref{contrac 1} with the specific terms involving $\delta_\xi \mathcal{B}$ and $\delta_\xi \overline{\mathcal{O}}$ extracted from \eqref{O2O3} yields
\begin{align}\label{left inte}
    \begin{aligned}
        &\int_{\mathcal{B}_L^>}\mathrm{d}^Dx\left[-(\delta_\xi\overline{\mathcal{O}})(\widehat{\Theta}_L^\mu)\partial_\mu\mathcal{B}+(\partial_\mu\overline{\mathcal{O}})(\widehat{\Theta}_L^\mu)\delta_\xi\mathcal{B}+(\delta_\xi\overline{\mathcal{O}})(\chi_L^\mu\mathcal{L})\partial_\mu\mathcal{B}-(\partial_\mu\overline{\mathcal{O}})(\chi_L^\mu\mathcal{L})\delta_\xi\mathcal{B}\right]\\
        &-\int_{\mathcal{B}_L^>}\mathrm{d}^{D}x(\partial_{\mu}\overline{\cal O})(\delta_{\xi}\mathcal{B}\widehat{\Theta}_{L}^{\mu})+\int_\mathcal{{\cal B}_L^>}\mathrm{d}^Dx(\delta_\xi\overline{\mathcal{O}})(\widehat{\Theta}_L^\mu)\partial_\mu\mathcal{B}\\
        =&\int_{{\cal B}_{{L}}^>} \mathrm{~d}^D x\left[\left(\delta_{\xi} \overline{\mathcal{O}}\right)\left(\chi_L^\mu \mathcal{L}\right) \partial_\mu \mathcal{B}-\left(\partial_\mu \overline{\mathcal{O}}\right)\left(\chi_L^\mu \mathcal{L}\right) \delta_{\xi} \mathcal{B}\right].
    \end{aligned}
\end{align}
Combining the first two terms in \eqref{contrac 1} with the final term in \eqref{Omega3}, and canceling the first term in the second line of \eqref{O2} against the $\delta\mathcal{B}$-dependent part of $Y_\mu$ in \eqref{contrac 1}, the remaining terms yield
\begin{align}\label{integrable charge 1}
    -\iota_{\xi}\left(\Omega_{(1)}+\Omega_{(2)}+\Omega_{(3)}\right)=\delta Q_{c}+\delta Q_{b}+\int_{{\cal B}_L^>}\left[\left(\delta_{\xi} \overline{\mathcal{O}}\right)\left(\mathcal{\chi_L^\mu{\cal L}}\right) \partial_\mu \mathcal{B}-\left(\partial_\mu \overline{\mathcal{O}}\right)\left({\chi_L^\mu{\cal L}}\right) \delta_{\xi} \mathcal{B}\right].
\end{align}

Furthermore, to evaluate the symplectic form associated with the boundary Lagrangian based on \eqref{ambi sym}, we apply the variational rule \eqref{variationrule}. We set $P_\epsilon(\mathcal{B})=\Delta(\mathcal{B})$, $N_\epsilon(\mathcal{C})=H(\mathcal{C})$, and $X=-\overline{\cal O}(V^\mu\partial_\mu{\cal B})$. Here, $V^\mu$ is evaluated using \eqref{dxl} and the Lie derivative $\mathcal{L}_{\chi}\ell^{\mu}=\chi^{\alpha}\partial_{\alpha}\ell^{\mu}-\ell^{\alpha}\partial_{\alpha}\chi^{\mu}+\ell^{\mu}\partial_{\alpha}\chi^{\alpha}$, yielding
\begin{align}
    V^\mu := \delta\ell^\mu + \chi_L^\mu\partial_\nu\ell^\nu = \partial_\alpha(2\ell^{[\alpha}\chi_{L}^{\mu]}) + \ell^\mu\partial_\alpha\chi_{L}^\alpha+{\cal A}^\mu.
    \end{align}
Consequently, we obtain the symplectic form induced by the boundary Lagrangian
\begin{align}\label{ellsym}
\Omega_\epsilon^{(\ell)}&=-\int_{\mathcal{B}_{L}^>}\mathrm{d}^{D-1}{x}\overline{\mathcal{O}}\left[\delta V^\mu+\partial_\alpha(\chi_L^\alpha\wedge V^\mu)\right]\partial_\mu{\cal B}-\int_{{\cal B}_L^>}[\delta{\overline{\cal O}} \wedge(V^\mu\partial_\mu{\cal B})+(\partial_{\alpha}\overline{\mathcal{O}})({\chi}_{L}^{\alpha}\wedge V^{\mu}\partial_{\mu}\mathcal{B})].
\end{align}
To further decompose this result into contributions localized on $\mathcal{S}_L$, we define a antisymmetric tensor $C_{\mathcal{S}}^{\alpha\mu}$ as follows
\begin{align}
C_{\mathcal{S}}^{\alpha\mu}:=2\chi_{L}^{[\alpha}\wedge\delta\ell^{\mu]}+\chi_{{L}}^{\alpha}\wedge\chi_{{L}}^{\mu}\partial_{\beta}\ell^{\beta}.
\end{align}
 This tensor satisfies 
\begin{align}
  \delta V^\mu+\chi_L^\alpha\wedge\partial_\alpha V^\mu-V^\alpha\wedge\partial_\alpha\chi_L^\mu-\partial_{\alpha}C_{\mathcal{S}}^{\alpha\mu}=\partial_\alpha\chi_L^\mu\wedge(2\delta\ell^\alpha+\chi_L^\alpha\partial_\nu\ell^\nu)-\partial_\alpha\chi_L^\alpha\wedge V^\mu.
\end{align}
This relation, when taken together with \eqref{ellsym} ultimately yields
\begin{align}
  - \int_{\mathcal{B}_{L}^>}\mathrm{d}^{D-1}{x}\overline{\mathcal{O}}\left[\delta V^\mu+\partial_\alpha(\chi_L^\alpha\wedge V^\mu)\right]\partial_\mu{\cal B}=\Omega_\epsilon^{(\ell{\cal B})}+\Omega_\epsilon^{(\ell{\cal S})},
\end{align}
where 
\begin{subequations}\label{OS boulk}
    \begin{align}
    \Omega_\epsilon^{(\ell{\cal B})}:=&-\int_{\mathcal{B}_{L}^{>}}\mathrm{d}^{D-1}x\overline{\mathcal{O}}_{T}^{L}\left[-(\partial_{\alpha}\chi_{L}^{\mu})\wedge V^{\alpha}+(\partial_{\alpha}\chi_{L}^{\mu})\wedge(2\delta\ell^{\alpha}+\chi_{L}^{\alpha}\partial_{\nu}\ell^{\nu})\right]\partial_{\mu}{\cal B}+\int_{\mathcal{B}_{\mathbf{L}}^{>}}\mathrm{d}^{\mathbf{D}-\mathbf{1}}{x}\left(\partial_{\alpha}\overline{\mathcal{O}}\right)({C}_{\mathcal{S}_L}^{\alpha\mu})\partial_{\mu}\mathcal{B}\notag\\
    &-\int_{\mathcal{S}_L} \mathcal{E}_{\mu \nu} \overline{\mathcal{O}} (V^\mu \chi_L^\nu),\\
        \Omega_\epsilon^{(\ell{\cal S})}:=&\frac{1}{2}\int_{\mathcal{S}_{L}}\mathrm{d}^{D-2}x\mathcal{E}_{\mu\nu}\overline{\mathcal{O}}_{T}^{L}C_{\cal S}^{\mu\nu}.
    \end{align}
\end{subequations}
Furthermore, utilizing the definition of $V^\mu$, the terms involving $\delta_\xi \overline{\mathcal{O}}$ and $\partial_\mu{\mathcal{O}}$ in \eqref{ellsym} can also be decomposed into contributions localized on $\mathcal{S}_L$ and $\mathcal{B}_L^>$:
\begin{subequations}
\begin{align}\label{dxiV}
&-\int_{\mathcal{B}_L^{\geq}}\delta\overline{\mathcal{O}}\wedge(V^\mu\partial_\mu\mathcal{B})=-\int_{\mathcal{B}_L}\mathrm{d}^{D-1}x\left[\delta\overline{\mathcal{O}}\wedge(\ell^\mu\partial_\alpha\chi_L^\alpha)-(\partial_\alpha\delta\overline{\mathcal{O}})\wedge(2\ell^{[\alpha}\chi_L^{\mu]})+\delta\overline{\mathcal{O}}\wedge\mathcal{A}^\mu\right]\partial_\mu{\cal B}\notag\\
&-\int_{\mathcal{S}_L}\mathrm{d}^{D-2}x\mathcal{E}_{\mu\nu}\delta\overline{\mathcal{O}}\wedge(\ell^{[\mu}\chi_L^{\nu]}),\\
&-\int_{\mathcal{B}_L^>}\mathrm{d}^{D-1}x(\partial_\alpha\overline{\mathcal{O}})\left[{\chi}_L^\alpha\wedge V^\mu\right]\partial_\mu\mathcal{B}=-\int_{\mathcal{S}_{\mathrm{L}}}\mathrm{d}^{{D}-2}{x}(\partial_{\alpha}\overline{\mathcal{O}})\left[{\chi}_{{L}}^{\alpha}\wedge{\ell}^{[\mu}{\chi}_{{L}}^{\nu]}\right]\mathcal{E}_{\mu\nu}\notag\\
&- \int_{\mathcal{B}_L^>} \mathrm{d}^{D-1}x \Bigg\{ 2 (\partial_\nu \partial_\alpha \overline{\mathcal{O}}) \Big[ \chi_L^\alpha \wedge \ell^{[\mu}\chi_{L}^{\nu]} \Big] + 2 (\partial_\alpha \overline{\mathcal{O}}) \Big[ (\partial_\nu \chi_L^\alpha) \wedge \ell^{[\mu}\chi_{L}^{\nu]} \Big] + (\partial_\alpha \overline{\mathcal{O}}) \Big[ \chi_L^\alpha \wedge \big( \ell^\mu\partial_\nu\chi_{L}^\nu + {\cal A}^\mu \big) \Big] \Bigg\} \partial_\mu \mathcal{B}.
\end{align}
\end{subequations}
It can be verified that the $\Omega_\epsilon^{(\ell{\cal S})}$ part and the integral over ${\cal S}_L$ in \eqref{dxiV} can be combined into an integrable charge. This is achieved using the method from \cite{Speranza:2022lxr,Chandrasekaran:2020wwn}, which introduces a corner ambiguity term into the dressed fields. In our case, we introduce a tensor $\beta^{\mu\nu}$ that shifts the symplectic form according to:
$\Omega_\epsilon \to \Omega_\epsilon + \delta\int_{{\cal S}_L}{\cal E}_{\mu\nu}\beta^{\mu\nu}$. Applying the variational rule prescribed in \eqref{z variation} yields the specific form of the corner ambiguity term employed here 
\begin{align}
    \beta^{\mu\nu}:=\overline{\mathcal{O}}(\ell^{[\mu}\chi_L^{\nu]}).
\end{align}
This ensures that certain bulk and corner terms in \eqref{OS boulk} and \eqref{dxiV} combine with the symplectic ambiguities to form an integrable term:
 \begin{align}\label{inte am}
    &-\iota_\xi \Omega_\epsilon^{(\ell{\cal S})}+\iota_\xi\int_{\mathcal{S}_L}\mathrm{d}^{D-2}x\mathcal{E}_{\mu\nu}\delta\overline{\mathcal{O}}\wedge(\ell^\mu\chi_L^\nu)+\iota_\xi \int_{\mathcal{S}_{\mathrm{L}}}\mathrm{d}^{{D}-2}{x}(\partial_{\alpha}\overline{\mathcal{O}})\left[{\chi}_{{L}}^{\alpha}\wedge{\ell}^{[\mu}{\chi}_{{L}}^{\nu]}\right]\mathcal{E}_{\mu\nu}+\iota_\xi\int_{\mathcal{S}_L} \mathcal{E}_{\mu \nu} \overline{\mathcal{O}} (V^\mu \chi_L^\nu)\notag\\
    &+\iota_\xi\delta\int_{{\cal S}_L}{\cal E}_{\mu\nu}\beta^{\mu\nu}=\delta\int_{\mathcal{S}_{L}}\mathrm{d}^{D-2}x\mathcal{E}_{\mu\nu}\overline{\cal O}\left(2\xi^{[\mu}\ell^{\nu]}\right),
 \end{align}
where we use the Cartan's rule in field space $\delta_\xi=\iota_{\xi}\delta+\delta\iota_{\xi}={\cal L}_\xi$.

By adding the terms on ${\cal B}_L^>$ in \eqref{integrable charge 1} to the remaining contractions in $-\iota_\xi \Omega_\epsilon^{(\ell)}$ (excluding those from \eqref{inte am}), we obtain the non-integrable symplectic contraction, $\slashed{\delta}Q_{NI}$:
\begin{align}\label{charge_and_flux}
\slashed{\delta}Q_{NI}=&\delta Q_{b}+\int_{{\cal B}_L^>}\left[\left(\delta_{\xi} \overline{\mathcal{O}}\right)\left(\mathcal{\chi_L^\mu{\cal L}}\right) \partial_\mu \mathcal{B}-\left(\partial_\mu \overline{\mathcal{O}}\right)\left({\chi_L^\mu{\cal L}}\right) \delta_{\xi} \mathcal{B}\right]\notag\\
&-\int_{\mathcal{B}_{L}^{>}}\mathrm{d}^{D-1}x\overline{\mathcal{O}}_{T}^{L}\left[-(\partial_{\alpha}\chi_{L}^{\mu})\wedge V^{\alpha}+(\partial_{\alpha}\chi_{L}^{\mu})\wedge(2\delta\ell^{\alpha}+\chi_{L}^{\alpha}\partial_{\nu}\ell^{\nu})\right]\partial_{\mu}{\cal B}\notag\\
&-\int_{\mathcal{B}_L^>}\mathrm{d}^{D-1}x\left[\delta\overline{\mathcal{O}}\wedge(\ell^\mu\partial_\alpha\chi_L^\alpha)-(\partial_\alpha\delta\overline{\mathcal{O}})\wedge(2\ell^{[\alpha}\chi_L^{\mu]})+\delta\overline{\mathcal{O}}\wedge\mathcal{A}^\mu\right]\partial_\mu{\cal B}+\int_{\mathcal{B}_{\mathbf{L}}^{>}}\mathrm{d}^{\mathbf{D}-\mathbf{1}}{x}\left(\partial_{\alpha}\overline{\mathcal{O}}\right)({C}_{\mathcal{S}}^{\alpha\mu})\partial_{\mu}\mathcal{B}\notag\\
=&\int_{\mathcal{B}_L^>}\mathrm{d}^{D-1}x\left\{(\partial_\alpha\overline{\mathcal{O}})\left[\mathcal{I}_1^{\alpha\mu}\partial_\mu{\cal B}\right]+(\delta\overline{\mathcal{O}})\left[\mathcal{I}_2^\mu\partial_\mu{\cal B}\right]-\overline{\mathcal{O}}\left[(\mathcal{I}_3^\mu-\mathcal{I}_4^\mu)\partial_\mu{\cal B}\right]\right\}\notag\\
=&\int_{\mathcal{B}_L^>}\mathrm{d}^{D-1}x\left\{\left[\partial_b(\mathcal{I}_1^{a\mu}\partial_\mu\mathcal{B})\right]({b_1})\partial_\alpha b_1+\left[\partial_b(\mathcal{I}_2^\mu\partial_\mu\mathcal{B})\right](b_2)\delta b_2-\left[(\mathcal{I}_3^\mu-\mathcal{I}_4^\mu)\partial_\mu\mathcal{B}\right](b_{3,4})\right\},
\end{align}
where the introduced definitions
\begin{subequations}
    \begin{align}
        {\cal I}_1^{\alpha\mu}:&=2V^\mu(\iota_\xi\chi_L^\alpha)-2\chi_L^\alpha(\iota_\xi V^\mu)-V^\mu\xi^\alpha+\xi^\mu\delta\ell^\alpha+2\xi^{[\alpha}\chi_L^{\mu]}\mathcal{L}-\delta Q_N^{\alpha\mu}(\xi)+\chi_L^\mu(\iota_\xi\delta\ell^\alpha),\\
        {\cal I}_2^\mu:=&\iota_\xi V^\mu+\partial_\alpha Q_N^{\alpha\mu}(\xi),~~~
        {\cal I}_3^\mu:=\partial_\alpha(\iota_\xi\chi_L^\mu)\delta\ell^\alpha,~~~
        {\cal I}_4^\mu:=(\partial_\alpha\chi_L^\mu)(\iota_\xi\delta\ell^\alpha)
    \end{align}
\end{subequations}
ensure that  $\overline{\mathcal{O}}\left[\mathcal{I}_{1}^{\alpha\mu}\partial_\mu\mathcal{B}\right]=\left[\mathcal{I}_1^{\alpha\mu}\partial_\mu\mathcal{B}\right]_{b=b_{1}}$ and $\overline{\mathcal{O}}\left[\mathcal{I}_{2,3,4}^\mu \partial_\mu \mathcal{B}\right]=\left[\mathcal{I}_{2,3,4}^\mu \partial_\mu \mathcal{B}\right]_{b=b_{2,3,4}}$ are satisfied under the condition in \eqref{trans_of_operator_mai}. According to \eqref{charge_and_flux}, the condition for the total symplectic contraction $-\iota_\xi \Omega_\epsilon$ to be integrable on $\mathcal{B}_L^>$ is
\begin{align}\label{integrble condition}
    \delta b_2=\frac{\left[\left(\mathcal{I}_3^\mu({b_{3}})-\mathcal{I}_4^\mu({b_{4}})\right) \partial_\mu \mathcal{B}\right]-\left[\partial_b\left(\mathcal{I}_1^{\alpha \mu}({b_1}) \partial_\mu \mathcal{B}\right)\right] \partial_\alpha b_1}{\left[\partial_b\left(\mathcal{I}_2^\mu(b_2) \partial_\nu \mathcal{B}\right)\right]}.
\end{align}
Condition \eqref{integrble condition} ensures integrability for the symplectic contraction of fields governed by displacement relations. This contrasts with the approach in \cite{Speranza:2019hkr,Chandrasekaran:2020wwn,Speranza:2022lxr}, where integrability relies on Dirichlet boundary conditions (equivalent to $\delta b_L = 0$ and $\delta\epsilon = 0$ in our framework). Here, the boundary dynamics on $\mathcal{B}$ are entirely encoded within the fields $b_{1,2,3,4}$. It can be verified that, according to the displacement relation in \eqref{shifted relation}, when both $b_L$ and $\epsilon$ are constants and $\delta b_2$, $\mathcal{I}_3^\mu-\mathcal{I}_4^\mu$ and $\partial_{\mu} b_1$ all vanish, relation \eqref{integrble condition} is automatically satisfied.

Finally, under condition \eqref{integrble condition}, the non-integrable part can be discarded, allowing us to identify the integrable charges in 
\begin{align}
    -\iota_{\xi}\Omega_{\epsilon}=\delta\int_{\mathcal{S}_{L}}\mathrm{d}^{D-2}x\mathcal{Q}_{I}(\xi),
\end{align}
where the integrand in the first term on the right-hand side above is simply the operator acting on the Noether charge plus the ambiguity correction from the boundary Lagrangian
\begin{align}\label{integrable_charge}
\mathcal{Q}_I(\xi) := \overline{\mathcal{O}} [ (-Q_N(\xi)+2\mathcal{E}_{\mu\nu}\xi^{[\mu}\ell^{\nu]} ) ] 
\end{align}
The integrability condition in \eqref{integrble condition}, derived by using the pointwise dependence property, is effectively an equation for $\delta b_L$ on $\mathcal{B}_L^>$, as can be seen from the definition of $b_{\mathcal{F}}$ in the displacement relation \eqref{shifted relation}. Generally, $\slashed{\delta}Q_{{NI}}$ represents boundary fluctuations and operator-induced fluxes, which cannot be neglected. However, the physical interpretation of the integrable charge $\mathcal{Q}_I(\xi)$ remains unclear. In the next section, we use a specific $\epsilon$ to illustrate the significance of this integrable charge.

It is noted that the displacement relation \eqref{shifted relation} is similar to that in \cite{Grumiller:2019fmp,Adami:2020ugu,Ruzziconi:2020wrb,Adami:2021sko,Adami:2021kvx,Adami:2020amw}, in which the choice of slicing of boundaries determines the field dependence of symmetry generators. Crucially, it renders surface charges integrable by genuine news from coordinate artifacts. Furthermore, the slicing dictates the specific structure of the boundary symmetry algebra, such as the Heisenberg algebra \cite{Afshar:2016wfy,Grumiller:2016kcp}. The difference, however, is that we place the newly introduced fields at the field-dependent singularities of the Dirac delta functions.

\section{Noether charge at the asymptotic boundary of general relativity in the FG gauge}
\label{sec:Noether charge at the asymptotic boundary of general relativity in the FG gauge}
We consider GR as an example to further analyze the integrable charge with soft cutoffs, specifically focusing on the case where the boundary is free of fluctuations and $\epsilon$ is a constant. The investigation of asymptotic symmetries in asymptotically locally AdS spacetimes necessitates a rigorous definition of the phase space \cite{Hawking:1982dh}. To tame the divergences inherent in the Einstein-Hilbert action, we employ the holographic renormalization procedure. Originally established to render the on-shell action finite, this method introduces local boundary counter-terms to subtract radial divergences appearing in the Fefferman-Graham (FG) gauge \cite{Emparan:1999pm,deHaro:2000vlm,Bianchi:2001kw,Skenderis:2002wp}.

The original action we consider including the Einstein-Hilbert action $S_{EH}$ and the Gibbons-Hawking term $S_{GH}$ is
 \begin{align}\label{GR action}
     S_{GR}=S_{EH}+S_{GH}=\frac{1}{2 \kappa^2} \int_{\cal M} {\rm d}^D x \sqrt{-g}L_{EH}+\frac{1}{\kappa^2} \int_{\partial{\cal M}} {\rm d}^{D-1} x \sqrt{-\gamma} K,
 \end{align}
where $L_{EH}=R-2 \Lambda$  and $\partial {\cal M}$ is the boundary of the entire manifold $\cal M$. For asymptotically $\rm{AdS}$, this includes both the asymptotic boundary and the horizon. In this action, $\kappa^2$ is equal to $8\pi G_N$, where $G_N$ is Newton's constant.  $R$ denotes Ricci scalar. On the boundary, $\gamma$ is the determinant of the induced metric $\gamma_{ij}$, while $K$ is the trace of the extrinsic curvature, introduced to cancel derivative terms from the bulk variation. For $\rm{AdS}$ space with negative curvature, the standard physical relation between the cosmological constant $\Lambda$ and the $\rm{AdS}$ radius $\ell$ is $\Lambda=-\frac{(D-1)(D-2)}{2 \ell^2}$. The equations of motion obtained by varying the metric in \eqref{GR action}, along with the adopted FG gauge, are
\begin{subequations}
\begin{align}
   R_{\mu\nu}&=-\frac{D-1}{\ell^{2}}g_{\mu\nu},\label{eom einstein}\\
  ds^2&=g_{\mu\nu}dx^\mu dx^\nu=\frac{\ell^2}{z^2}dz^2+\gamma_{ij}(\bar x,z){\rm d}\bar x^i{\rm d}\bar x^j,
\end{align}
\end{subequations}
where $i,j=1,\cdots, D-1$ and we recall that $\bar{x}$ represents the non-radial coordinates. We identify the radial coordinate $z$ in the FG gauge with the coordinate $b$ in \eqref{extended action}. The expansion of the induced  metric and its determinant near the asymptotic boundary $z=0$ are
\begin{subequations}\label{FG expansion}
\begin{align}
    \gamma_{ij}(\bar x,z)&=\frac{\ell^2}{z^{2}}\left[g_{(0)ij}+z^{2}g_{(2)ij}+\cdots+z^{D-1}g_{(D-1)ij}+{h_{(D-1)ij}z^{D-1}}\ln{z^{2}}+\ldots\right],\label{FG expansion1}\\
    \sqrt{-\gamma}&=\sqrt{g_{(0)}}\left(\frac{\ell}{z}\right)^{(D-1)}\left[1+\frac{1}{2}z^2T_2+\cdots+z^{D-1}\left(\frac{1}{2}T_{D-1}+\frac{1}{2}T_h\ln z^2\right)\right],\label{FG expansion2}
\end{align}
\end{subequations}
where logarithmic terms appear only when the boundary dimension $D-1$ is even \cite{deHaro:2000vlm}. Here, we define $T_n=\mathrm{Tr}(g_{(0)}^{-1}g_{(n)})$ and $T_h=\mathrm{Tr}(g_{(0)}^{-1}h_{(D-1)})$, with the trace $\mathrm{Tr}$ taken by contracting indices with the metric $g^{(0)ij}$. We identify $z$ with the $b$ coordinate in \eqref{extended action}. This implies that, to identify the corrections induced by the T- and S-type operators, we must first obtain the asymptotic behavior of the Noether charge with respect to $r$. The trace of the extrinsic curvature is $K= \gamma^{\mu\nu}\nabla_\mu n_\nu$ with outward unit normal $n^\mu=-\frac{z}{\ell}\delta^\mu_z$. To this end, we adopt the method of \cite{Papadimitriou:2005ii} and define $\lambda$ such that the on-shell Lagrangian satisfies $\frac{1}{2\kappa^2}\sqrt{-g}{L}_{EH}\simeq\sqrt { - g} \frac{{1 - D}}{{{\kappa ^2}{\ell ^2}}}=-\frac{1}{\kappa^2}\frac{\rm d}{{\rm d}z}\left(\sqrt{-\gamma}\lambda(z)\right)$. Then the alternative form of $L_{EH}$ and the equation satisfied by $\lambda$ can be obtained:
\begin{subequations}
\begin{align}
& S_{EH}=\frac{1}{\kappa^2}\int_{z=\varepsilon}{\rm d}^{D-1}x\sqrt{-\gamma}\lambda,\\
&\frac{z}{\ell}\partial_z\lambda-K\lambda+\frac{1}{2}{L}_{EH}=0\label{lambda equation},
\end{align}
\end{subequations}
where $\varepsilon$ is a cutoff defined by $z=\varepsilon$ near the asymptotic boundary. Then the action in \eqref{GR action} can be cast into the form
\begin{align}\label{Gr action}
   S_{GR}= \frac{1}{\kappa^2}\int_{z=\varepsilon}{\rm d}^{D-1}x\left[\sqrt{-\gamma}\sum_{n=0}^\infty(K-\lambda)_{(n)}\right],
\end{align}
where we only focus on the term of the action at $z=\varepsilon$. Based on the definition of the extrinsic curvature and equation \eqref{lambda equation}, we can verify that $K$, ${ L}_{EH}$ and $\lambda$ can be expanded in the forms\footnote{These expansions are derived via the identities $\det M=e^{\mathrm{Tr}\ln M}$ and $\partial_z(\ln\gamma)=(\gamma^{ij}\partial_z\gamma_{ij})$, $(I+M)^{-1}=I-M+M^2-\ldots$, where $I$ is an identity matrix and $M$ is an arbitrary invertible matrix. Since the majority of these results were previously demonstrated in \cite{deHaro:2000vlm}, we omit the details here.}:
\begin{subequations}\label{KLL expansion|}
\begin{align}
    K &= K_{(0)} + z^2K_{(2)} + \cdots + K_{D-1}z^{D-1} + K_h z^{D-1}\ln z^2, \\
    \sqrt{-g} L_{\text{on-shell}} &= -2(D-1)\ell^{D-2}\sqrt{g_{(0)}}\left[z^{-D} + \frac{1}{2}T_{2}z^{-D+2} + \cdots + z^{-1}\left(\frac{1}{2}T_{D-1} + \frac{1}{2}T_{h}\ln z^{2}\right)\right], \\
    \lambda(z) &= \lambda_{(0)} + \lambda_{(2)}z^2 + \cdots + z^{D-1}\lambda_{(D-1)} + z^{D-1}\ln z^2\widetilde{\lambda}_{(D-1)},
\end{align}
\end{subequations}
where $K_h$ arises from the logarithmic term in \eqref{FG expansion1}. Based on \eqref{lambda equation}, $\sqrt{-\gamma}(K-\lambda)$ yields the non-logarithmic terms
\begin{align}
 \sum_{n=0,2,\cdots}(K-\lambda)_{(n)}\propto \sum_{n=0,2,\cdots}{z^{n-D+1}}.
\end{align}
We find that as $\varepsilon \to 0$, terms with $n < D-1$ diverge and require cancellation via $I_{ct}$, the $n = D-1$ term yields a finite result, while terms with $n > D-1$ vanish. Therefore, to cancel this divergence, an additional counterterm
\begin{align}\label{counter term}
    S_{ct}=-\frac{1}{\kappa^{2}}\int_{z=\varepsilon}{\rm d}^{D-1}x\sqrt{-\gamma}(K_{(n)}-\lambda_{(n)})_{ct},
\end{align}
with $(K_{(n)}-\lambda_{(n)})_{ct}:=\sum_{n=0}^{D-2}(K-\lambda)_{(n)}+(\widetilde{K}_{(D-1)}-\widetilde{\lambda}_{(D-1)})\ln z^2$ is required. Thus, the renormalized action is
\begin{align}
  S_{ren}=S_{GR}+S_{ct}.
\end{align}

Identifying the time coordinate $t$ with the previously used $c$, the vector field $\xi = \partial_t$ generates time translations. Its contraction $i_\xi$ maps the boundary volume element $\mathrm{d}^{D-1} \bar{x}$ to $\mathrm{d}^{D-2}\hat{x}\sqrt{\sigma}$, with $\sigma$ denoting the determinant of the induced metric $\sigma_{\mu\nu}$ on $\mathcal{S}$. We therefore directly adopt the Noether charge formulation from \cite{Papadimitriou:2005ii}, which allows us to directly read off the finite Noether charges from the renormalized action $S_{\mathrm{ren}}$:
\begin{align}
    Q_\xi=\int_{\mathcal{S}}{\mathrm{d}^{D-2}\hat{x}\sqrt{\sigma}}({\cal Q}_N[\xi]-\mathbf{B}_\xi),
\end{align}
where
\begin{subequations}\label{total in charge}
\begin{align}
{\cal Q}_N(\xi)&=-\frac{1}{\kappa^2}\sqrt{\sigma}n_\mu u_\nu\nabla^\mu\xi^\nu,\\
    \int_{\mathcal{S}}\mathbf {B}_\xi&=\int_{\mathcal{S}}\mathrm{d}^{D-2}\hat{x}\sqrt{\sigma}\frac{u_\mu\xi^\mu}{\kappa^2}[K-(K-\lambda)_{ct}].
\end{align}
\end{subequations}
Here, $u_\mu = -1/\sqrt{-\gamma^{tt}}\nabla_\mu t$ denotes the timelike unit normal vector, which implies that the binormal takes the form $2n_{[\mu} u_{\nu]}$. The term $\mathbf {B}_\xi$ originates from contracting the integrand in \eqref{counter term} with the vector field $\xi$, in combination with the relation $\left.i_\xi(\sqrt{-\gamma}\mathrm{d}^{D-1}\bar x)\right|_\mathcal{S}=(-u_\mu\xi^\mu)\sqrt\sigma\mathrm{d}^{D-2}\hat{x}$.  We provide the first two orders of $K$ and $\lambda$ as examples:
\begin{subequations}\label{KLL expansion|2}
    \begin{align}
        K_{(0)}&=\frac{D-1}{\ell},\quad K_{(2)}=-\frac{1}{\ell}\mathrm{Tr}(g_{(2)})=\frac{\ell}{2(D-2)}{\cal R},\\
         \lambda_{(0)}&=\frac{1}{\ell},\quad \lambda_{(2)}=-\frac{1}{D-3}K_{(2)},
    \end{align}
\end{subequations}
in which $\cal R$ corresponds to the contraction of the Riemann tensor associated with the metric $g^{(0) ij}$: ${\cal R}=g^{(0) ij}{\cal R}_{(0)ij}$ and the calculation of $K_{(2)}$ refers to the relation $\mathrm{Tr}(g_{(2)})=-\frac{\ell^2}{2(D-2)}{\cal R}$.

To proceed, we extend the action \eqref{GR action} to a form comparable to \eqref{FG expansion}:
\begin{align}
  \mathcal{L}=\frac{\sqrt{-g}L_{EH}}{2\kappa^2}{\rm d}^ D x,\quad \quad \ell^\mu=\frac{\sqrt{-g}n^\mu K}{\kappa^2} {\rm d}^ {D-1} \bar x.
\end{align}
We shall demonstrate that upon implementing the extension of the action, the resulting integrable charge $\mathcal{Q}_I$ in \eqref{integrable_charge} is equivalent to the charge derived from the action \eqref{GR action} supplemented with the counterterm in \eqref{counter term}. Consequently, the following relation holds
\begin{align}\label{key corrected relation}
   \widetilde{{\cal O}}{\cal Q}_{I}=-{\cal Q}_{ct},
\end{align}
where ${\cal Q}_{I}$ and ${\cal Q}_{ct}$ can be obtained from \eqref{total in charge} with the subleading operator $ \widetilde{{\cal O}}$ of $\overline{{\cal O}}$, respectively. Their explicit forms are given by
\begin{subequations}
    \begin{align}
         \widetilde{{\cal O}}:=&\sum_{n=1}^\infty\frac{(-1)^n M_n}{n!}\partial_z^n,\\
         {\cal Q}_I(z,\bar x)=&{\cal Q}_N(\xi)-\sqrt{\sigma}\frac{u_\mu\xi^\mu}{\kappa^2}K=\frac{\sqrt{-\gamma}}{\kappa^2}\left(K_i^t\xi^i+K\xi^t\right),\\
        \mathcal{Q}_{ct}(z,\bar x)=&\frac{\sqrt{-\gamma}}{\kappa^2}\left[\sum_{n=0}^{D-2}(K-\lambda)_{(n)}+(\widetilde{K}_{(D-1)}-\widetilde{\lambda}_{(D-1)})\ln z^2\right]\label{Qct}.
    \end{align}
\end{subequations}
In ${\cal Q}_I$, the equalities  $-\sqrt{\sigma}\frac{u_\mu\xi^\mu}{\kappa^2}K=\frac{\sqrt{-\gamma}}{\kappa^2}K\xi^t$ and $\nabla^z\xi^t=-\frac{z}{\ell}K_i^t\xi^i$ have been used. It can be verified that, under the expansions \eqref{FG expansion}, \eqref{KLL expansion|} and \eqref{KLL expansion|2}, ${\cal Q}_N$ and ${\cal Q}_{ct}$ take the following expansion forms
\begin{subequations}\label{cal Q expansion}
    \begin{align}
        \mathcal{Q}_I(z,\bar x)&=\frac{\sqrt{-g_{(0)}}}{\kappa^2\ell}\left[\sum_{k=0}^\infty I_k(\bar x)z^{-(D-1-k)}+\ln z^2\sum_{k=D-1}^\infty\tilde{I}_k(\bar x)z^{-(D-1-k)}\right],\\
        \mathcal{Q}_{ct}(z,\bar x)&=\frac{\sqrt{-g_{(0)}}}{\kappa^2\ell}\left[\sum_{k=0}^\infty a_k(\bar x)z^{-(D-1-k)}+\ln z^2\sum_{k=D-1}^\infty\tilde{a}_k(\bar x)z^{-(D-1-k)}\right].
    \end{align}
\end{subequations}

To investigate the physical significance of the operator-valued integrable charge, we shall now utilize $I_k$, $\tilde I_k$, $a_k$, and $\tilde a_k$ in combination with \eqref{key corrected relation} to solve $M_n$. To this end, we effectively restrict our analysis to even spacetime dimensions $D$ and consider $\tilde I_k$ and $\tilde a_k$ to vanish for simplicity. Given that \eqref{Qct} contains exactly $D-1$ terms, we must impose a truncation on $M_n$ such that all terms beyond $M_{D-1}$ identically vanish. Consequently, an additional set of $D-1$ equations is required to satisfy \eqref{key corrected relation}.  Given $\partial_z^n(z^{-p})=(-1)^n\frac{(p+n-1)!}{(p-1)!}z^{-p-n}$, we first introduce the relation
\begin{align}
    \widetilde{\mathcal{O}}(z^{-p})=\sum_{n=1}^{D-1}\frac{m_n}{n!}(-z)^n\left[(-1)^nn!\binom{p+n-1}{n}z^{-p-n}\right]=P(p)z^{-p},
\end{align}
with the convention for the binormal coefficent $\binom{n}{k}=\frac{n!}{k!(n-k)!}$ and $P(p) := \sum_{n=1}^\infty m_n \binom{p+n-1}{n}$, where the coefficients $m_n$ and $M_n$ are linked by the relation $M_n = m_n z^n$. Then it follows that the relation between $I_k$ and $a_k$ in \eqref{cal Q expansion} is
\begin{align}
    I_kP(D-1-k)=-a_k,
\end{align}
which is derived from $\widetilde{\mathcal{O}}\left(z^{-(D-1-k)}\right)=P(D-1-k)z^{-(D-1-k)}$. Then we can use this result to solve for $m_n$ as
\begin{align}\label{final mn}
{m_n = \sum_{w=n}^N \sum_{j=1}^w (-1)^{n+j+1} \binom{w-1}{n-1} \binom{w}{j} \frac{a_{D-1-j}}{I_{D-1-j}}},
\end{align}
with $N$ the truncation limit of the series, which is detailed in Appendix \ref{app:Solutions for moments in holographic renormalization}. By treating the parameters $\varepsilon$ and $\epsilon$ strictly as constants, we obtain the solution
\begin{align}
M_n = m_n z^n \quad \text{with}\quad z=\varepsilon=\epsilon.
\end{align}
This not only confirms the validity of \eqref{key corrected relation} but also provides a concrete physical meaning for the charge in \eqref{integrable_charge}. As demonstrated by example \eqref{example_G}, the integral $M_n=(-1)^n\epsilon^n\int_{-\infty}^\infty u^n\frac{{\rm d}G(u,\bar x)}{{\rm d}u}{\rm d}u$ depends on both $\epsilon$ and the integral of $F(u,\bar x)$. This dependence ensures the self-consistency of result \eqref{final mn}, successfully aligning the ratio $a_{D-1-j}/I_{D-1-j}$ with $M_n$. Ultimately, this provides a precise criterion for interpreting the expansion coefficients in \eqref{OT operator}, wherein the DRF is utilized to guarantee finiteness at the asymptotic boundary.

Note that in an asymptotically AdS background, traditional background subtraction is replaced by holographic renormalization. Infinite divergences from the bulk action are removed by adding local boundary counterterms \cite{Papadimitriou:2005ii}. These terms depend solely on intrinsic boundary geometry and act as  universal subtractor terms \cite{Balasubramanian:1999re}. This mechanism renders the variational problem well-posed without requiring a specific reference spacetime. Consequently, it yields finite conserved charges defined via the renormalized stress-energy tensor ensuring physical consistency across different topologies \cite{Henningson:1998gx}. Our results demonstrate that DRFs can be employed to implement a soft-cutoff procedure, which naturally yields finite renormalized Noether charges and identifies $\epsilon$ as the radial boundary.

\section{Conclusions and discussions}
\label{sec:conclusions}
In this paper, we generalize covariant actions from hard to soft cutoffs via DRFs, treating the smearing function as a small boundary thickness field. Imposing fluctuating boundary conditions on this function yields the extended covariant Lagrangian in \eqref{extended action} and restricted MCFs in \eqref{MC constarint}. Consequently, this smearing effectively shifts the spacetime boundaries outward by the thickness scale, generating the field-dependent relational boundaries shown in \eqref{relational_action}.

By selecting DRFs that implement a soft cutoff, we find that these DRFs are characterized by an infinite-order expansion of the Dirac delta function with respect to the thickness field, arising from the T-type and S-type functions introduced in  \eqref{OT operator} and \eqref{OS operator}.  The MCFs depend not only on the fluctuating boundaries; specifically, the smearing function is required to satisfy the boundary conditions shown in below  \eqref{difference action}. They are also influenced by the DRFs, which naturally emerge from the variational principle of the Lagrangian in relational coordinates. Our results show that the MCF acts as a compensation term for the coordinate transformation induced by extending the boundary outward \eqref{MCF in component}, which defines how differential forms transform at the boundaries. Furthermore, the MCF is constrained by boundary conditions. In the absence of boundary fluctuations, its form is restricted by equation \eqref{MC constarint}, and it becomes equivalent to the linear MCF $\chi_{L,R}$ under the fluctuating boundary conditions specified in \eqref{new boundary con}.

Furthermore, we investigate the covariant phase space formalism in the presence of soft cutoffs for the boundary subsystem. We found that the MCF describes the fluctuating boundaries and the thickness field, appearing in both the symplectic potential and the symplectic form, as shown in \eqref{left right symplectic potential} and \eqref{main symplectic form}. Crucially, by introducing the singularities of the field-dependent Dirac delta function via the displacement relation \eqref{shifted relation}, we can cancel the non-integrable components of the surface charges. The thickness field thus ensures charge integrability at new field-dependent locations, yielding the integrable charges \eqref{integrable_charge} associated with the redefined operator $\overline{\mathcal{O}}_L^T$. In this case, introducing the boundary Lagrangian allows the integrability condition \eqref{integrble condition} under fluctuating boundary conditions to serve as a constraint governing the boundary dynamics.

These studies offer a perspective on the renormalization of Noether charges. Specifically, DRFs can be employed to extend covariant theory actions to a soft-cutoff formulation. By selecting a specific profile for the Dirac delta function, we ensure the finiteness of the Noether charges at the asymptotic boundary, as demonstrated in \eqref{final mn}.

Building on the soft cutoff method adopted in this paper, we can identify several intriguing directions for future research. First, we could clarify the underlying gravitational subsystems and their symmetries in more detail. For instance, considering the null hypersurfaces and finite causal diamonds discussed in \cite{Chandrasekaran:2026pnc}, it remains to be seen whether the pullback of the soft-cutoff symplectic potential to the hypersurface allows for a standard decomposition into a boundary term, a corner term, and a flux \cite{Harlow:2019yfa,Chandrasekaran:2018aop}. Additionally, replacing the Heaviside step function used in hard cutoffs for symmetry generators with our smearing functions would make it possible to compute the integrable charges associated with supertranslations. A second interesting direction is to extend the soft cutoff method to other theories, such as Maxwell theory. Building upon the work in \cite{Kabel:2023jve}, we can investigate whether the soft cutoff formalism can accommodate both diffeomorphism and $\mathrm{U}(1)$ symmetries, along with their corresponding quantum reference frames.

Moreover, for the FG gauge considered in this paper, one can apply the field-dependent transformation proposed in \cite{Ciambelli:2024vhy} to transition to the Bondi gauge. In this case, we can explore the relationship between the thickness field and residual diffeomorphisms to realize integrable charges for the field-dependent $\epsilon$ under restricted boundary conditions.

\section*{Acknowledgment}
This work is supported by the Natural Science Foundation of China under Grants No.~12375054, the Postgraduate Research $\&$ Practice Innovation Program of Jiangsu Province under Grants No.~KYCX-3502, the Natural Science Foundation of Sichuan Province under Grants No.~2026NSFSC0745 and the Doctoral Research Initiation Project of China West Normal University under Grants No.~25KE038.

\appendix
\section{Proofs for equations regarding DRFs}
\label{app:Proofs for equations regarding DRFs}
In this appendix, we provide a general proof of \eqref{variation rule} and the explicit form of the result in \eqref{variation UL}.
Let $V = U^{-1}$ denote the inverse map of $U$ from $\mathfrak{m}$ to $\mathcal{M}$, given by $x^\mu = V^\mu(y)$, and the spacetime components of the $p$-form field $\alpha$ be given by $\alpha_{\mu_1\dots\mu_p}(x)$. The components of its pushforward $U^\star \alpha$ on the relational spacetime $\mathfrak{m}$ are defined as:
\begin{align}
	(U^\star\alpha)_{A_1\cdots A_p}(y)=\alpha_{\mu_1\cdots\mu_p}(V(y))\prod_{k=1}^pJ_{A_k}^{\mu_k}(y),
\end{align}
where the Jacobi matrix is defined as $J_A^\mu(y)=\frac{\partial V^\mu}{\partial y^A}$. According to the Leibniz rule, we give the expression of the left hand side of \eqref{variation rule} to get
\begin{align}\label{check of variation rule}
	\begin{aligned}
	\delta(U^\star\alpha)& =\delta\left[\alpha_{\mu_1\cdots\mu_p}(V)\prod_{k=1}^pJ_{A_k}^{\mu_k}\right] =\delta[\alpha_{\mu_1\cdots\mu_p}(V)]\left(\prod_{k=1}^pJ_{A_k}^{\mu_k}\right)+\alpha_{\mu_1\cdots\mu_p}(V)\delta\left(\prod_{k=1}^pJ_{A_k}^{\mu_k}\right).
\end{aligned}
\end{align}
 The variation $\delta$ acts on both the explicit form of the field $\alpha$ and the map $V$:
\begin{align}\label{first term of check of variation rule}
	\begin{aligned}
		\delta[\alpha_{\mu_1\cdots\mu_p}(V(y))] & =(\delta\alpha)_{\mu_1\cdots\mu_p}|_{x=V(y)}+\frac{\partial\alpha_{\mu_1\cdots\mu_p}}{\partial x^\nu}\delta V^\nu(y)=(\delta\alpha)_{\mu_1\cdots\mu_p}+(\partial_\nu\alpha_{\mu_1\cdots\mu_p})\chi^\nu.
	\end{aligned}
\end{align}
By imposing the condition $[\delta, \partial_y] = 0$, we compute the action of $\delta$ on $J_A^\mu$ as
\begin{align}
	\delta J_A^\mu=\delta\left(\frac{\partial V^\mu}{\partial y^A}\right)=\frac{\partial}{\partial y^A}(\delta V^\mu)=\frac{\partial}{\partial y^A}(\chi^\mu(V(y)))=(\partial_\nu\chi^{\mu})J_{A}^\nu.
\end{align}
Inserting the above relation into the second term on the RHS of the second line of \eqref{check of variation rule} gives
\begin{align}\label{second term of check of variation rule}
\begin{aligned}
\alpha_{\mu_{1}\cdots\mu_{p}}(V)\delta\prod_{k=1}^{p}J_{A_{k}}^{\mu_{k}} & =\alpha_{\mu_1\cdots\mu_p}(V)\sum_{m=1}^p(\partial_\nu\chi^{\mu_m})J_{A_m}^\nu\prod_{k\neq m}J_{A_k}^{\mu_k}\\
 & =\sum_{m=1}^p\left[\alpha_{\mu_1\cdots\mu_m\cdots\mu_p}(\partial_\nu\chi^{\mu_m})\right]J_{A_1}^{\mu_1}\cdots J_{A_m}^\nu\cdots J_{A_p}^{\mu_p}
\end{aligned}.
\end{align}

For the right-hand side of \eqref{variation rule}, using the definition of covariant variation given by $\delta_\chi\alpha=\delta\alpha+\mathcal{L}_\chi\alpha$, i.e.,
\begin{align}
	(\mathcal{L}_\chi\alpha)_{\rho_1\cdots\rho_p}=\chi^\nu\partial_\nu\alpha_{\rho_1\cdots\rho_p}+\sum_{m=1}^p\alpha_{\rho_1\cdots\lambda\cdots\rho_p}(\partial_{\rho_m}\chi^\lambda),
\end{align}
we obtain
\begin{align}
	\begin{aligned}\label{third term of check of variation rule}
		[U^\star(\delta_\chi \alpha)]_{A_1\cdots A_p}& =[U^\star(\delta\alpha+\mathcal{L}_\chi\alpha)]_{A_1\cdots A_p},\\
		&={\left[(\delta\alpha)_{\rho_1\cdots\rho_p}+\chi^\nu\partial_\nu\alpha_{\rho_1\cdots\rho_p}\right]\prod_{k=1}^pJ_{A_k}^{\rho_k}} +\left[\sum_{m=1}^p\alpha_{\rho_1\cdots\lambda\cdots\rho_p}(\partial_{\rho_m}\chi^\lambda)\right]\prod_{k=1}^pJ_{A_k}^{\rho_k}.
	\end{aligned}
\end{align}
Collecting all these, it is straightforward to prove the validity of \eqref{variation rule}.

Next, we express the variational relation \eqref{variation UL} in component form to clarify some formal conventions. The variation of the Lagrangian is the volune element. The pushforward of $\cal L$ can be written as
\begin{align}
    (U^\star {\cal L})=L\left(x(y)\right)\det\left(\frac{\partial x^\mu}{\partial y^A}\right){\rm d}y^1\wedge{\cdots\wedge {\rm d}y^D.}
\end{align}
With the Jacobian determinant $J=\det\left(\frac{\partial x^\mu}{\partial y^A}\right)$, the variation of $U^\star {\cal L}$ takes the form
\begin{align}
    \delta(U^\star {\cal L})=\left(\delta[L(x)] J+L\delta J\right){\rm d}^Dy.
\end{align}
Further, the variations $\delta(L(x))$ and $\delta J$ are
\begin{subequations}
\begin{align}
    &\delta [L(x(U))]=\delta L+\chi^\mu\partial_\mu L,\\
    & \delta J=J\frac{\partial y^A}{\partial x^\mu}\delta\left(\frac{\partial x^\mu}{\partial y^A}\right)=J\partial_\mu\chi^\mu,
\end{align}
\end{subequations}
where we have incorporated the MCF $\delta x^\mu=\chi^\mu$ and the variation of the Jacobian determinant given by $\delta J=J\mathrm{tr}[(\frac{\partial x}{\partial y})^{-1}\delta\frac{\partial x}{\partial y}]$ with $\mathrm{tr}$ denoting the trace operation.  Then we have
\begin{align}
    \delta(U^\star {\cal L})=J\left[\delta L+\left(\chi^\mu\partial_\mu L+L\partial_\mu\chi^\mu\right)\right]{\rm d}^Dy=U^\star(\delta {\cal L} + {{\cal L}_\chi }{\cal L}).
\end{align}

Using Cartan's magic formula: ${\cal L}_{\xi} \bullet ={\rm d}i_\xi\bullet+i_\xi{\rm d}\bullet$ for arbitrary vector $\xi$, the Lie derivative of $\mathcal{L}$ along the direction $\chi=\chi^\mu\frac{\partial}{\partial x^\mu}$ can be rewritten as
\begin{align}
	\left.\mathcal{L}_\chi {\cal L}=	{\rm d}(i_\chi{ \cal L}\right)+ i_\chi	{\rm d} {\cal L}.
\end{align}

Since $\mathcal{L}$ is a $D$-form on the $D$-dimensional spacetime, so we have ${\rm d}{\cal L}=0$, and  $\delta{}{\cal L}=E_\Phi[\Phi]\delta\Phi+{\rm d}{\Theta_{LW}}[\Phi,\delta\Phi]$ with
\begin{align}\label{LW sym p}
{\Theta_{LW}}[\Phi,\delta\Phi]=\Theta_{LW}^\mu[\Phi,\delta\Phi]{\rm d}^{D-1}x_{\mu}.
\end{align}
Further combining
\begin{subequations}
	\begin{align}
		 i_\chi {\cal L}&={L} \chi^\mu{\rm d}^{D-1} x_\mu,\\
		 {\rm d}(i_\chi {\cal L})&={\rm d}[{L} \chi^\mu{\rm d}^{D-1} x_\mu]=\partial_\nu\left({L} \chi^\mu\right) {\rm d} x^\nu \wedge{\rm d}^{D-1} x_\mu,
	\end{align}
\end{subequations}
shall give us an explicit form of \eqref{variation UL}.

\section{Calculation of expansion coefficients}
\label{app:Calculation of Expansion Coefficients}
In this appendix, we detail the derivation of the expansions in \eqref{main expansion two} and \eqref{LS operator relation}, along with the expressions for $M_n^{L,R}$ and $W_n^{L,R}$.  By applying the chain rule to calculate the partial derivatives of ${\cal H}_b$ in \eqref{smearing function}, we obtain
\begin{align}\label{He}
\frac{{\rm d}\mathcal{H}_{b}}{{\rm d}\epsilon}=-\left[\frac{u_{L}}{\epsilon}\frac{\partial G_{L}}{\partial u_L}\right]+\left[\frac{u_{R}}{\epsilon}\frac{\partial G_{R}}{\partial u_R}\right].
\end{align}
Bsed on this, let $\Psi(b)$ be an arbitrary smooth test function. We consider an integral of the following form
\begin{align}
  I_1=\int_{b_0}^b\left[\frac{u_{L,R}}{\epsilon}\frac{\partial G(u_{L,R})}{\partial u_{L,R}}\right]\Psi(b){\rm d}b.
\end{align}
  Expanding $\Psi(b)$ in a Taylor series around the boundaries $b=b_{L,R}$, we get
  \begin{align}
    \Psi(b)=\Psi(b_{L,R}+\epsilon u_{L,R})=\sum_{n=0}^\infty\frac{\epsilon^nu_{L,R}^n}{n!}\Psi^{(n)}(b_{L,R}),
  \end{align}
and plug this expansion into $I_1$, we obtain
\begin{align}
    I_1=\sum_{n=0}^\infty\frac{\epsilon^n}{n!}\Psi^{(n)}(b_{L,R})\int_{-\infty}^\infty u_{L,R}^{n+1}\frac{\partial G(u_{L,R})}{\partial u_{L,R}}{\rm d}u_{L,R},
\end{align}
where for the second equality, we considered the relations ${\rm d}b=\epsilon{\rm d}u_{L,R}$ at the left and right boundaries, respectively. For the right-hand side of \eqref{main expansion one}, we have
\begin{align}
    I_2=\int_{-\infty}^\infty\left[\sum_{n=0}^\infty\frac{W_n^{L,R}}{n!}\Delta^{(n)}(b-b_{L,R})\right]\Psi(b){\rm d}b.
\end{align}
 Using the property $\int\delta^{(n)}(b)\Psi(b)=(-1)^n\Psi^{(n)}(0)$ and set $I_1=I_2$, we get
 \begin{align}\label{W sol}
    W_n^{L,R}=(-1)^n\epsilon^{n-1}\int_{b_0}^b u_{L,R}^{n+1}\frac{\partial G(u_{L,R})}{\partial u_{L,R}}{\rm d}b=(-1)^n\epsilon^n\int_{u_0}^{u} u_{L,R}^{n+1}\frac{\partial G(u_{L,R})}{\partial u_{L,R}}{\rm d}u_{L,R}.
 \end{align}
  Similar steps can be used to derive the expression for $M_n^{L,R}$. Then, further recalling  \eqref{He},  we prove \eqref{main expansion two}.

 $M_{n+1}^{L,R}$ can be expressed by $W_{n}^{L,R}$ as
\begin{align}\label{WM relation}
    \begin{aligned}
        \begin{aligned}
M_{n+1}^{L,R} & =(-1)^{n+1}\epsilon^{n+1}\int_{-\infty}^{\infty}{u_{L,R}}^{n+1}\frac{\partial G_{L,R}}{\partial u_{L,R}}\partial u_{L,R}=-\epsilon\left[(-1)^n\epsilon^n\int_{-\infty}^{\infty}u_{L,R}^{n+1}\frac{\partial G_{L,R}}{\partial u_{L,R}}\partial u_{L,R}\right]=-\epsilon W_n^{L,R},
\end{aligned}
    \end{aligned}
\end{align}
in which we set the integration limits to $\pm\infty$, ensuring that $u_{L,R}$ spans the entire real line, which in turn implies $-\infty < u_{L,R} < \infty$ as $\epsilon\to0$. Subsequently, we find that the left-hand side of  \eqref{LS operator relation} can be written as
\begin{equation}
\begin{split}
   & (b-b_{L,R})\sum_{n=0}^\infty\frac{M_n^{L,R}}{n!}\partial_b^n\Delta(b-b_{L,R})
    = -\sum_{n=1}^\infty\frac{M_n^{L,R}}{(n-1)!}\partial_b^{n-1}\Delta(b-b_{L,R})\\
    &= \epsilon\sum_{k=0}^\infty\frac{W_k^{L,R}}{k!}\partial_b^k\Delta(b-b_{L,R})= \epsilon {\cal O}_S\Delta(b-b_{L,R}).
\end{split}
\end{equation}
Thus, \eqref{LS operator relation} is proved.

\section{Solutions for $m_n$ in holographic renormalization}
\label{app:Solutions for moments in holographic renormalization}

In this appendix, we derive the solution for $m_n$ in \eqref{final mn}. We define the difference operator $\delta_t := E - I$, where $E f(s) := f(s+1)$ and $I f(s) := f(s)$ for an arbitrary discrete function $f$.  By choosing $f=\binom{s+n-1}{n}$, we obtain
\begin{align}
   \delta_t\binom{s+n-1}{n}=\binom{s+n}{n}-\binom{s+n-1}{n}=\binom{s+n-1}{n-1}.
\end{align}
Then, we obtain the action of $\delta_t$ raised to the power of $w$
\begin{align}
    \delta_t^w\binom{s+n-1}{n}=\binom{s+n-1}{n-w}.
\end{align}
Based on this property and the definition of $P(s)$, the action of $\delta_t^w$ on $P(s)$ yields
\begin{align}
  \delta_t^w P(s)=\sum_{k=1}^N m_k\delta_t^w\binom{s+k-1}{k}=\sum_{k=1}^N m_x\binom{s+k-1}{k-w},
\end{align}
where $N=D-1$. Setting $s=0$ and given that $\binom{k-1}{k-w} = 0$ is vanishing for all $w<k$, we have
\begin{align}\label{p1}
    \delta_t^wP(0)=\sum_{k=1}^Nm_k\binom{k-1}{k-w}=\sum_{k=w}^Nm_k\binom{k-1}{w-1}.
\end{align}
Additionally, result $\delta_t^wP(s)=\sum_{j=0}^w(-1)^{w-j}\binom{w}{j}E^jP(s)=\sum_{j=0}^{w}(-1)^{w-j}\binom{w}{j}P(s+j)$ gives us
\begin{align}\label{p2}
    \delta_t^wP(0)=\sum_{j=0}^w(-1)^{w-j}\binom{w}{j}P(j)=\sum_{j=0}^w(-1)^{w-j}\binom{w}{j}P(j)=\sum_{j=0}^w(-1)^{w-j}\binom{w}{j}V_{D-1-j},
\end{align}
where we further introduce $P(D-1-k)=-\frac{a_k}{I_k}:=V_k$ follows from the definition of $P(s)$. Comparing \eqref{p1} and \eqref{p2}, an equation of $m_k$ is given
\begin{align}
  \sum_{k=w}^Nm_k\binom{k-1}{w-1}  =\sum_{j=1}^w(-1)^{w-j}\binom{w}{j}V_{D-1-j}.
\end{align}
Multiplying both sides by $(-1)^{w-n} \binom{w-1}{n-1}$ and summing over $w$ from $n$ to $N$ one can isolate $m_k$. To do so, we use the identities
\begin{align}
 \binom{k-1}{w-1}\binom{w-1}{n-1}=\binom{k-1}{n-1}\binom{k-n}{w-n},\quad\quad  \sum_{l=0}^{k-n}(-1)^l\binom{k-n}{l}=(1-1)^{k-n}.
\end{align}
Applying these relations, we obtain
\begin{align}
    m_k=\sum_{w=k}^N(-1)^{w-k}\binom{w-1}{k-1}\left[\sum_{j=1}^w(-1)^{w-j}\binom{w}{j} V_{D-1-j}\right],
\end{align}
which gives \eqref{final mn}.

\bibliography{Ref}
\bibliographystyle{utphys}


\end{document}